\pdfoutput=1
\documentclass[12pt]{article}
\usepackage{jheppub}
\usepackage{amsfonts}
\usepackage{mathrsfs}
\usepackage{amsmath}
\usepackage{amssymb}
\usepackage{bm}
\usepackage[normalem]{ulem}
\usepackage{extarrows}
\usepackage{slashed}
\usepackage{isodateo}
\usepackage{graphicx}
\graphicspath{{fig/}}

\usepackage{xcolor}
\usepackage{indentfirst}
\newcommand{\FR}[2]{\displaystyle\frac{\,{#1}\,}{#2}}
\newcommand{\fr}[2]{\mbox{$\frac{\,{#1}\,}{#2}$}}
\newcommand{\n}{\nonumber}
\renewcommand{\rm}{\mathrm}

\def\bge{\begin{equation}}
\def\ede{\end{equation}}
\def\bga{\begin{aligned}}
\def\eda{\end{aligned}}
\def\bgp{\begin{pmatrix}}
\def\edp{\end{pmatrix}}
\def\bgs{\begin{subequations}}
\def\eds{\end{subequations}}
\newcommand{\order}[1]{\mathcal{O}({#1})}
\def\di{{\mathrm{d}}}
\def\D{{\mathrm{D}}}

\def\mb{\mathbf}

\def\pd{\partial}

\def\la{\langle}\def\ra{\rangle}
\def\sla{\slashed}

\def\tr{\mathrm{\,tr\,}}

\def\to{\rightarrow}
\def\To{\Rightarrow}
\def\ii{\mathrm{i}}

\def\al{\alpha}
\def\be{\beta}
\def\ga{\gamma}
\def\de{\delta}
\def\ep{\epsilon}

\def\lam{\lambda}
\def\rh{\rho}
\def\si{\sigma}

\def\I{\mathcal{I}}

\def\Mp{M_{\text{Pl}}}

\def\OSM{\mathcal{O}_{\text{SM}}}
\def\O{{\mathcal{O}}}

\newcommand{\ob}[1]{\mkern 2mu \overline{\mkern -2mu #1 \mkern -2mu}\mkern 2mu}
\newcommand{\wt}[1]{\mkern 2mu \widetilde{\mkern -2mu #1 \mkern -2mu}\mkern 2mu}
\newcommand{\wh}[1]{\mkern 2mu \widehat{\mkern-2mu#1\mkern-2mu}\mkern 2mu}

\title{Standard Model Mass Spectrum in Inflationary Universe}

\author[a]{Xingang Chen,}
\author[b]{Yi Wang,}
\author[c]{and Zhong-Zhi Xianyu}

\affiliation[a]{Institute for Theory and Computation, Harvard-Smithsonian Center for Astrophysics,\\
60 Garden Street, Cambridge, MA 02138, USA}

\affiliation[b]{Department of Physics, The Hong Kong University of Science and Technology,\\
Clear Water Bay, Kowloon, Hong Kong, P.R.China}

\affiliation[c]{Center of Mathematical Sciences and Applications, Harvard University,\\
20 Garden Street, Cambridge, MA 02138, USA}

\emailAdd{xingang.chen@cfa.harvard.edu}
\emailAdd{phyw@ust.hk}
\emailAdd{xianyu@cmsa.fas.harvard.edu}

\abstract{
We work out the Standard Model (SM) mass spectrum during inflation with quantum corrections, and explore its observable consequences in the squeezed limit of non-Gaussianity. Both non-Higgs and Higgs inflation models are studied in detail. We also illustrate how some inflationary loop diagrams can be computed neatly by Wick-rotating the inflation background to Euclidean signature and by dimensional regularization.
}

\begin{document}
\maketitle

{

\section{Introduction}

Recent development in the inflationary perturbation theory has revealed a new way of probing some of the highest energy particle states in our universe. All particles present in the inflationary universe with mass up to the Hubble scale $H$ leave characteristic signals in soft limits of primordial non-Gaussianities. Remarkably, these signals directly encode the mass and spin spectrum of these particles \cite{Chen:2009we,Chen:2009zp,Baumann:2011nk,Assassi:2012zq, Noumi:2012vr, Arkani-Hamed:2015bza, Chen:2015lza,Lee:2016vti, Sefusatti:2012ye,Norena:2012yi,Meerburg:2016zdz,
Chen:2012ge,Gong:2013sma, Kehagias:2015jha,Dimastrogiovanni:2015pla,Schmidt:2015xka, Emami:2013lma,Bonga:2015urq}, making primordial perturbations a particle detector of the early universe.

As in ground-based colliders, before exploring new physics, we have to understand the signals from the Standard Model (SM) of particle physics. If the Hubble scale of inflation is much larger than the electroweak broken scale, one might na\"ively treat all SM fields as being effectively massless and ignore them if our goal is to explore the much heavier states accessible to the cosmological collider.
However, the situation is more complicated -- Through loop corrections, some light fields can acquire large mass with the inflationary background. In \cite{Chen:2016nrs}, it is shown that a classically massless scalar boson can receive nonzero mass due to its self interaction. Gauge boson can also receive similar mass correction if there exist some light scalar particles charged under the corresponding gauge group. On the contrary, classically massless fermions do not receive nonzero Dirac mass correction from its Yukawa interaction with light scalars. This result may also be understood qualitatively from the point of view of the mean field approximation. The expectation value of Higgs-field-squared is of order $H^2/\sqrt{\lam}$ for the massless case (where $\lam$ is the self-coupling of Higgs field), due to the Gibbons-Hawking temperature $H/2\pi$ of the inflationary background, even though the vacuum expectation value (VEV) of the Higgs field is zero. This provides the origin of the masses of certain fields.

In this paper, we continue this line of research and work out the SM particle spectrum in inflation models and their imprints in the primordial non-Gaussianities.

In Sec.\;\ref{Sec_2pt}, we revisit 1-loop mass corrections to various particles. A similar calculation is done in \cite{Chen:2016nrs} using real-time Schwinger-Kelydish formalism and in a particular space-time asymmetric gauge. The nonzero mass corrections there result from resumming the infrared-divergent loop diagrams by the dynamical renormalization group (DRG) method \cite{Burge09}. The calculation has a number of subtleties and is technically involved, too. Given both technical and conceptual importance of the loop correction to SM spectrum, in Sec.\;\ref{Sec_2pt} of this paper, we shall present an alternative derivation of the same result by carrying out all loop calculations in Euclidean de Sitter (dS) space. Euclidean dS approach has the advantage that the full spacetime symmetry is manifest and is made good use of. With some tricks, the loop calculation can also be done easily and neatly. Some useful tools for doing calculation in Euclidean dS are collected in Appendix \ref{AppEdS} of this paper.

In Sec.\;\ref{Sec_mass}, Sec.\;\ref{Sec_3pt}, and Sec.\;\ref{Sec_Higgs}, we study the mass spectrum of SM in inflation models and their signatures in bispectrum.
Apart from loop corrections, the interactions between inflaton and SM fields can also introduce nonzero mass correction to SM fields, due to the nonzero inflaton background. The inflaton-SM couplings may depend heavily on inflation models.
In Sec.\;\ref{Sec_mass} and Sec.\;\ref{Sec_3pt}, we study generic non-Higgs inflation models in which the Higgs field has zero VEV.
On the other hand, if the inflaton itself \textit{is} the SM Higgs boson \cite{Bezrukov:2007ep,Bezrukov:2013fka} -- a class of models known collectively as Higgs inflation -- the inflaton-SM coupling would be very similar, though not identical, to various Higgs couplings in SM. We study the case of Higgs inflation separately in Sec.\;\ref{Sec_Higgs}.

The main lesson we learn from the analysis of SM is that many particles in SM can well receive mass of $\order{H}$. It is then natural to study the signals of SM fields in primordial bispectrum following \cite{Chen:2009we,Chen:2009zp,Baumann:2011nk,Assassi:2012zq, Noumi:2012vr, Arkani-Hamed:2015bza}. The purpose of this study is twofold: On one hand, if the SM fields do acquire mass $\order{H}$ and do generate observable signals in the bispectrum, these imprints would constitute the background signal to the Cosmological Collider, of which we should have good understanding before using the Cosmological Collider to explore new physics at very high scales. On the other hand, the study of 1-loop SM contribution to the bispectrum with all spin-$(0,1/2,1)$ particles can serve as a prototypical example which can be readily generalized to other similar calculations with new physics included. In Sec.\;\ref{Sec_3pt} of this paper, we shall present detailed calculation of the squeezed limit of bispectrum contributed by all kinds of SM fields through 1-loop, in a generic non-Higgs inflation model with approximate shift symmetry for the inflaton. Meanwhile, we shall also discuss a parallel calculation in Higgs inflation in Sec.\;\ref{Sec_Higgs}. More physical consequences of this study was discussed in \cite{Chen:2016uwp}.

When mediated by a massive particle through tree diagram, the bispectrum has an angular dependence $P_s(\cos\theta)$ (where $\theta$ is the angle between the long and short mode and $P_s$ is the Legendre polynomial) that can tell us the spin $s$ of the particle \cite{Arkani-Hamed:2015bza}. However, since SM particles can only appear in loops in primordial non-Gaussianities, we expect some subtleties in the determination of the spin. The angular dependence only shows the total angular momentum of the loop, rather than the spin of an individual particle. More details can be found in Sec.\;\ref{Sec_3pt}.

On the observational side, the sensitivity of probing the primordial non-Gaussianities has been improving steadily. There has been a 300-fold improvement from the COBE era \cite{Komatsu:2001rj} to the Planck era \cite{Ade:2013ydc} in the past two decades. Future experiments on large-scale structure \cite{Amendola:2016saw,Dore:2014cca,Abell:2009aa} would further improve the precision by another order of magnitude. In the more distant future, the 21~cm experiments \cite{Furlanetto:2006jb, Pritchard:2011xb, Munoz:2015eqa, Meerburg:2016zdz} can potentially open up an enormous amount of observable volume and drastically reduce the cosmic variance, further improving the precision by a few orders of magnitude. While the experimental precision varies significantly for different types of bispectra, overall these
experiments are expected to be able to constrain the primordial bispectrum down to $f_{NL}\sim1$ for all major shapes of bispectra, and may even probe non-Gaussianities well below this value.
At $f_{NL} < 1$, for the local bispectrum, the density fluctuation from the curvaton and the inflaton may be distinguished \cite{Sasaki:2006kq}; for the equilateral bispectrum, we may tell whether the inflation mechanism is dominated by linear or nonlinear effects \cite{Creminelli:2003iq, Wang:2014kqa, He:2016uiy}.
The type of bispectra relevant to the cosmological collider physics is much more difficult to constrain than these conventional bispectra. Nonetheless, analysis show that the 21cm survey has the potential to probe these signals down to $f_{NL} > 10^{-2}$ \cite{Meerburg:2016zdz}.

Ideally, a detailed understanding of the SM background provides valuable information on what we could hope to learn from the experiments that probe the cosmological collider physics. In the parameter space where the SM background is observable, an agreement between our calculation and the observations would indicate a particle desert beyond SM up to the Hubble scale of inflation. On the other hand, observations that do not agree with the SM background of any parameter space would indicate new physics beyond the SM, such as new interactions or new particles. It would be interesting to work out the consequence of new physics, such as the GUT/supersymmetry/string states, in this setup.

We end this introductory section by a few remarks on the notations and conventions. The universe during inflation experiences nearly exponential expansion, with nearly constant Hubble parameter, and negligible spatial curvature. Such a spacetime can be well approximated by the Poincar\'e patch of the de Sitter space, which describes an exactly exponentially expanding universe with zero spatial curvature, with the following metric,
\bge
\label{dSFRW}
  \di s^2=-\di t^2+e^{2H t}\di \mb x^2,
\ede
where the comoving time $t\in(-\infty,\infty)$. The constant $t$ slices are flat and are parameterized by the comoving coordinates $\mb x$. The Hubble parameter $H$ is a real constant over time, and the exponential expansion is manifest through the scale factor $a^2(t)=e^{2Ht}$. The metric (\ref{dSFRW}) is conformally flat, and this can be seen by introducing the conformal time $\tau$ via $\di\tau^2=e^{-2Ht}\di t^2$. As a result, the metric (\ref{dSFRW}) becomes,
\bge
\label{dSc}
  \di s^2=\FR{1}{(H\tau)^2}(-\di\tau^2+\di\mb x^2),
\ede
where the conformal time $\tau\in(-\infty, 0)$, and it is convenient to fix the normalization by $-1/H\tau=e^{Ht}$. In this paper, we shall mostly work with conformal coordinates with metric (\ref{dSc}). To apply dimensional regularization, we shall sometimes work in $D=(d+1)$-dimensional dS, but eventually we shall take $D=4$ ($d=3$) limit of the result. In this paper we shall use both spacetime dimension $D$ and spatial dimension $d=D-1$ extensively.

\section{1-Loop Mass Correction Revisited}
\label{Sec_2pt}

In this section, we shall review the 1-loop correction to the masses of spin-$(0,1/2,1)$ particles. The loop correction in dS is important, due to the peculiar infrared behavior of scalar field. To see this in a simple way, we note that a minimally coupled massless scalar field $\phi$ in dS has a constant mode in the late time limit $\tau\to 0$. When the scalar is canonically normalized, its mode function in 3-momentum space is given by,
\bge
\label{phimode}
  \phi(\tau,\mb k)=\FR{H}{\sqrt{2k^3}}(1+\ii k\tau)e^{-\ii k\tau},
\ede
which indeed becomes a constant $\phi(0,\mb k)=H/\sqrt{2k^3}$ when $\tau\to 0$. Intuitively, when another particle $\chi$ interacts with $\phi$ through a time like $\lam\chi^2\phi^2$ in the Lagrangian, this constant mode can contribute a nonzero mass to $\chi$ field which is proportional to $\lam\la\phi^2\ra$ in the late-time limit $\tau\to 0$.

However, the expectation value $\la\phi^2\ra$, or more generally the 2-point function $\la\phi(x)\phi(x')\ra$, is ill-defined for a minimally coupled massless scalar field $\phi$, precisely because the infrared divergence coming from the constant zero mode. One can see this problem by noticing that the inverse-Fourier transformation of $\la\phi(\tau,\mb k)\phi(\tau,-\mb k)\ra$ back to coordinate space is ill-defined for the massless mode function (\ref{phimode}). It is also instructive to view this problem by Wick rotating the dS spacetime $dS_D$ to its Euclidean counterpart, which is simply a $D$-dimensional sphere $S^D$. On a sphere $S^D$, the scalar field $\phi$ can be decomposed into modes by spherical harmonics $Y_{\vec L}(x)$, where $\vec L=(L_D,L_{D-1},\cdots,L_2,L_1)$ is a $D$-dimensional vector taking values in integers and with the restriction $L_D\geq L_{D-1}\geq\cdots\geq L_2\geq|L_1|$ (More details are presented in Appendix \ref{AppEdS}). Throughout this paper we shall also denote the first entry $L_D$ by $L$. The constant mode in dS then corresponds to the zero mode $L=0$ on the sphere. In this setup, the tree-level 2-point function for a minimally coupled scalar field $\phi$ of mass $m$ is given by,
\bge
  \la \phi(x)\phi(x')\ra=\sum_{\vec L}\FR{1}{\lam_L}Y_{\vec L}^{}(x)Y_{\vec L}^*(x'),
\ede
where $\lam_L=L(L+d)+(m/H)^2$. Now it is clear that the zero-mode component of the 2-point function $1/\lam_0$ is divergent if $m=0$.

The divergence in zero mode is irrelevant for a free field $\phi$ as it is unobservable. It becomes important only when we turn on some coupling among fields so that zero modes interact with others. However, the appearance of a problem also provides a hint of the solution. The point here is that the zero mode
gets non-perturbatively coupled even when we turn on a small coupling.
As a result, the 1-loop calculation is insufficient and we must take account of a whole series of higher order loops. After summing over all these loops, a finite answer is obtained, and precisely has the form of a mass correction.

In \cite{Chen:2016nrs} the 1-loop corrections to 2-point functions of spin-$(0,1/2,1)$ fields are studied with the real time Schwinger-Keldysh formalism. The dynamical renormalization group resummation is used to sum over an important class of higher order loop diagrams. It is shown there that loop corrections can introduce nonzero mass to classically massless fields through infrared effects, especially if the scalar field in the loop has mass of order $H$ or less.

However, the results presented in \cite{Chen:2016nrs} also have some unwanted features which we would like to clarify. Firstly, the loop diagrams with 3-point vertices are calculated in such a way that both the time integral and momentum integral are artificially cut off at UV, and appear to be more divergent than loop diagrams with 4-point vertex in the IR. Although such results agree with similar calculations in literature \cite{Burge09,Meulen07}, they are nevertheless quite obscure. The second problem is that the in-in calculation treats the space and time separately so the manifest covariance of the results is lost. This explains why the time-time and space-space components of 2-point function for vector field have different behaviors at late times.

In this section we shall take another approach to this problem which makes the spacetime symmetry manifest and is also much simpler. This method involves the Wick rotation of time direction and does analysis in Euclidean version of de Sitter space. We present some basic material of Euclidean dS calculations in Appendix \ref{AppEdS}, where we also fix the notations for the following calculations. Readers interested in the calculation of this section may want to read Appendix \ref{AppEdS} before going on. The relation between in-in amplitudes in $dS_D$ and corresponding amplitudes in $S^D$ is carefully studied in \cite{Higuchi11} and we adopt the viewpoint that the two approaches are equivalent for the calculation we are interested in.

Before we proceed into the details, here we recall some general features of loop corrections presented in \cite{Chen:2016nrs}. Firstly, we can ignore all diagrams which do not contain scalar lines. Conceptually this is because the action for a gauge theory with charged massless fermions is classically Weyl invariant, and technically this is related to the fact that the mode functions for both gauge boson and fermion have no IR growth, so they do not contribute the IR divergence which is the source of the mass correction.

Secondly, we can also disregard the fermion-loop and vector-loop corrections to scalar's 2-point function. Once again, the technical reason is that such diagrams have no IR divergence. While this conclusion can be checked explicitly as was done in \cite{Chen:2016nrs}, one can also understand it by recalling the fact that the diagram with external massless scalars can be got by acting an appropriate differential operator (with respect to external momenta) on a corresponding diagram with all external massless scalars replaced by conformal scalars \cite{Arkani-Hamed:2015bza}. An example of this manipulation is given in (\ref{O12}) in Sec.\;\ref{Sec_3pt} of this paper. Now that there is no IR divergence in diagrams with conformal scalars, massless fermions, and vector fields, so we conclude that the diagrams with external scalars (but no loop scalars) do not contribute to mass correction.

Therefore, it only remains to consider diagrams with scalar loops. Now we are going to reevaluate 1-(scalar) loop corrections to 2-point functions of scalar, spinor, and vector fields, respectively, working in Euclidean dS space. The calculation can be very complicated if one demands full loop correction. However, we can simplify the calculation significantly by considering the mass correction only. Because the mass correction is independent of external momenta, we are free to set external fields to constants. Then the calculation becomes rather straightforward.

\subsection{A Toy Example}
\label{sec_toy}

To illustrate the basic technique of our calculation, it would be helpful to consider a toy example with two minimally coupled massive real scalar fields $\phi$ and $\chi$ interacting through a non-derivative cubic vertex. In $dS_D$, the action can be written as,
\bge
S_{\text{toy}}=-\FR{1}{2}\int\di^Dx\,\sqrt{-g}\Big[(\pd_\mu\phi)^2+(\pd_\mu\chi)^2+M_\phi^2\phi^2+M_\chi^2\chi^2+\lam\phi\chi^2\Big].
\ede
We would like to find the 1-loop correction to $M_\phi^2$ through $\chi$-loop, i.e. the external-momentum-independent piece of 1-loop correction to the 2-point function of $\phi$. For this purpose we work with $S^D$ and it is enough to set external $\phi$'s to be constants\footnote{To obtain the full loop correction to the propagator, one still need to work out the wave function renormalization, which can be got only by keeping external momentum finite. However, it can be easily understood that there is no infrared problem for this part, and the wave function renormalization will be essentially the same with the flat space counterpart. Therefore, we shall not consider wave function renormalization in this work, as its effect is expected to be subleading.}. Then we only need to evaluate the following integral,
\bge
\label{phi3loop}
  \FR{\lam^2\mu_R^{4-D}}{2}\int\di\Omega\di\Omega'\phi(x)\phi(x')G_{\chi}(x,x')^2,
\ede
where $\di\Omega=\di^Dx\sqrt{g(x)}$, $\di\Omega'=\di^Dx'\sqrt{g(x')}$ are invariant integral measures on $S^D$ at point $x$ and $x'$, respectively; $G_{\chi}(x,x')$ is the propagator for $\chi$, and a renormalization scale $\mu_R$ is introduced to keep the coupling $\lam$ being dimension 1 on $D$-sphere. In this expression we write $\phi(x)$ and $\phi(x')$ formally as two operators sitting at $x$ and $x'$ respectively, although it should be clear that both of them are constants. Then it is easy to work out the integral (\ref{phi3loop}) with the help of (\ref{ScaProSH}),
\begin{align}
\label{phi3loopRed}
&~\FR{\lam^2\mu_R^{4-D}H^{2D-4}}{2}\int\di\Omega\di\Omega'\,\sum_{\vec L,\vec M}\FR{1}{\lam_L\lam_M}Y_{\vec L}^{}(x)Y_{\vec L}^*(x')Y_{\vec M}^*(x)Y_{\vec M}^{}(x')\n\\
=&~\FR{\lam^2\mu_R^{4-D}H^{D-4}}{2}\int\di\Omega\, \sum_{\vec L}\FR{1}{\lam_L^2}Y_{\vec L}^{}(x)Y_{\vec L}^*(x)\n\\
=&~-\FR{\lam^2\mu_R^{4-D}H^{D-2}}{2}\FR{\pd}{\pd m^2}\int\di\Omega\,\sum_{\vec L}\FR{1}{\lam_L}Y_{\vec L}^{}(x)Y_{\vec L}^*(x)\n\\
=&~-\FR{\lam^2\mu_R^{4-D}}{2}\big[\pd_{m^2}G(x,x)\big]_{m^2=M_\chi^2}\int\di\Omega,
\end{align}
where $\lam_L=(L+d/2+\mu)(L+d/2-\mu)$ and $\mu=\sqrt{(d/2)^2-(m/H)^2}$, and we have used the orthonormal condition of spherical harmonics, and the fact that $G(x,x)$ is coordinate independent. According to (\ref{ScaPro}), the scalar propagator at coincident limit of its two variables is,
\bge
  G(x,x)=\FR{H^{D-2}}{(4\pi)^{D/2}}\FR{\Gamma(d/2-\mu)\Gamma(d/2+\mu)}{\Gamma(d/2)}\;{}_2F_1\Big(\FR{d}{2}-\mu,\FR{d}{2}+\mu;\FR{D}{2};1\Big).
\ede
where ${}_2F_1(a,b;c;z)$ is the hypergeometric function of type-$(2,1)$. Alternatively, we can also evaluate $G(x,x)$ from its mode decomposition (\ref{ScaProSH}) by carrying out the mode summation over $\vec L$ directly,
\begin{align}
\label{Lsum1}
  \sum_{\vec L}\FR{1}{\lam_L}=&~\sum_{L_D=0}^\infty\sum_{L_{D-1}=0}^{L_D}\cdots\sum_{L_2=0}^{L_3}\sum_{L_1=-L_2}^{L_2}\FR{1}{\lam_{L_D}}=\sum_{L_D=0}^\infty\FR{1}{\lam_L}\FR{(2L_D+d)\Gamma(L_D+d)}{\Gamma(d+1)\Gamma(L_D+1)}\n\\
  =&~\FR{d\Gamma(-d)\Gamma(d/2)\sin(\pi d/2)\cos(\pi\mu)}{\pi^{d/2}\Gamma(1-\frac{d}{2}-\mu)\Gamma(1-\frac{d}{2}+\mu)\big[\cos(\pi d)-\cos(2\pi\mu)\big]}.
\end{align}
The coincident limit is given by $G(x,x)=H^{-2}V_D^{-1}\sum\lam_L^{-1}$, where $V_D=2\pi^{(D+1)/2}/\Gamma(\frac{D+1}{2})\times H^{-D}$ is the volume of $S^D$. Then we can simplify $G(x,x)$ into the following form,
\bge
\label{Gxx}
  G(x,x)=\FR{H^{D-2}}{4(4\pi)^{D/2-1}}\FR{\Gamma(\mu+\fr{d}{2})\big[\tan(\fr{\pi d}{2})-\cot(\pi\mu-\fr{\pi d}{2})\big]}{\Gamma(\mu-\fr{d}{2}+1)\Gamma(\frac{1+d}{2})}.
\ede
This method of mode summation is particularly useful in the following when we calculate photon's 2-point function.

Now we are ready to evaluate the above result (\ref{phi3loopRed}) at $d=3-\epsilon$,
\begin{align}
&-\FR{\lam^2\mu_R^{4-D}}{2}\big[\pd_{m^2}G(x,x)\big]_{m^2=M_\chi^2}=\FR{\lam^2}{32\pi^2}\Big(\FR{2}{\ep}-\ga_E+\log4\pi\Big)-\FR{\lam^2}{32\pi^2\mu}\Big[(1-\mu+2\mu\log\FR{H}{\mu_R})\n\\
&~+2\mu\psi(\mu-1/2)+(1-\FR{M_\chi^2}{2H^2})\big(2\psi'(\mu-1/2)-\pi^2\sec^2(\pi\mu)\big)-\pi\mu\tan(\pi\mu)\Big]+\order{\ep},
\end{align}
where $\psi(z)$ is the digamma function. The divergent piece as $\ep\to 0$ is identical to the case of flat space, as it should. Now we use modified minimal subtraction ($\ob{\text{MS}}$) scheme to subtract the term proportional to $2/\ep-\ga_E+\log4\pi$, and then send $\ep=0$, to get the following 1-loop correction to $M_\phi^2$,
\begin{align}
  \de M_\phi^2=&-\FR{\lam^2}{16\pi^2\mu}\Big[(1-\mu+2\mu\log\FR{H}{\mu_R})+2\mu\psi(\mu-1/2)\n\\
&~+(1-\FR{M_\chi^2}{2H^2})\Big(2\psi'(\mu-1/2)-\pi^2\sec^2(\pi\mu)\Big)-\pi\mu\tan(\pi\mu)\Big].
\end{align}
In the case of inflation, we can set the renormalization scale $\mu_R=H$. It is interesting to note that the loop correction diverges as $M_\chi\to 0$,
\bge
  \de M_\phi^2=\FR{3\lam^2H^4}{8\pi^2M_\chi^4}+\order{M_\chi^0},
\ede
which should be expected. At the same time, the mass correction above is independent of $M_\phi$ and remains valid even when $M_\phi\to 0$. So we conclude that $\phi^3$-interaction can contribute a nonzero mass correction to classically massless scalar. The simplification of above calculation can be visualized as follows,
\bge
\parbox{0.6\textwidth}{\vspace{-5mm}\includegraphics[width=0.55\textwidth]{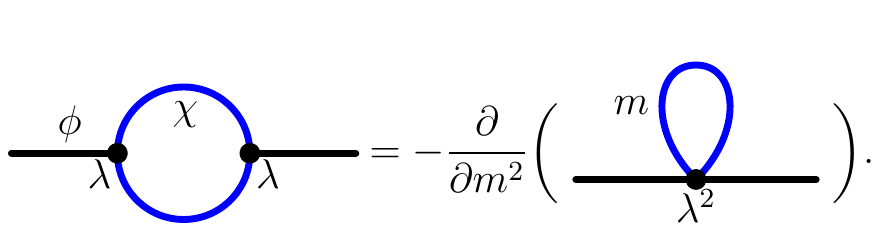}}
\ede
The left hand side of this expression is the original Feynman diagram in (\ref{phi3loop}), while the right hand side represents the final result of (\ref{phi3loopRed}) after our simplification.
It should be noted that the above diagrammatic expression is valid only when external black lines carry zero momentum.

\subsection{Loop Correction to Higgs Mass}
\label{SecZM}

For SM, we can treat Higgs field as massless field effectively so long as the inflation scale is much higher than the electroweak scale.
However, the Higgs sector of SM has quartic self-interaction instead of cubic, and needs a separate treatment which is quite different from the above toy model. 

For free scalar fields the zero mode is unobservable, and therefore, the divergence is physically irrelevant. On the other hand, once we turn interactions on, e.g. the quartic potential of Higgs field, an effective mass term for Higgs can be built dynamically and the divergence of zero modes is thus removed. In \cite{Chen:2016nrs} this is explored in the real time in-in calculation with dynamical renormalization group resummation. While this treatment can provide a reasonable qualitative description of the mechanism of dynamical mass generation which agrees with other methods such as stochastic approach, Large $N$ limit, and Euclidean method \cite{Burge09}, it is still not fully satisfactory since the DRG resummation cannot take account of all soft loop diagrams. In this respect, the Euclidean approach is again advantageous because the zero mode loops can be summed to all orders in perturbation theory \cite{Raja2010,Nacir2016}. Below we review this calculation very briefly.

The Higgs action can be written as,
\begin{align}
\label{SHiggs}
  S_{\text{Higgs}}=&-\int\di^Dx\sqrt{-g}\Big[|\D_\mu\mb H|^2+\lam (\mb H^\dag \mb H)^2\Big]\n\\
  \supset& - \int\di^Dx\,\sqrt{-g}\Big[\FR{1}{2}(\pd_\mu h_i)^2-\FR{1}{4}\lam(h_i h_i)^2\Big],
\end{align}
where $h_i~(i=1,\cdots, 4)$ denotes four real components of Higgs field. In terms of standard parameterization $\mb H=\frac{1}{\sqrt 2}(\pi_1+\ii\pi_2,h+\ii\pi_0)^T$ we may identify $(h_1,\cdots h_4)=(h,\pi_0,\pi_1,\pi_2)$. If we are allowed to decouple gauge boson by turning off the gauge couplings, then the Higgs sector would be consist of 4 real scalars with $O(4)$ symmetry, as is shown in the expression above. On the other hand, if the SM gauge symmetry is broken, either by the standard Higgs mechanism in flat space, or by the nontrivial zero modes in dS as will be shown below, the $(\pi_0,\pi_1,\pi_2)$ components will be unphysical, and only the Higgs boson $h$ remains in the physical spectrum.

In perturbation theory the tree-level propagator of the Higgs field is ill-defined due to the absence of the mass term and the divergence of the zero mode. The insight here, however, is that the zero modes can be treated nonperturbatively. In a general $\phi^4$ theory with tree level mass $m_0$ and with $O(\mathcal{N})$ symmetry, the 2-point function of the zero modes can be written as $\la h_i h_j\ra =\mathcal{N}^{-1}\la h^2\ra \delta_{ij}$, where $i,j=1,\cdots \mathcal{N}$, $h^2\equiv h_i h_i$ is the square of the radial direction in the field space, and the expectation value $\la\cdots\ra$ is taken with respect to the Bunch-Davis vacuum in Euclidean dS. The expectation value $\la h^2\ra$ can be evaluated directly as follows,
\begin{align}
\label{h2vev}
\la h^2\ra\equiv&~\FR{\int\di^\mathcal{N}h\,h^2\exp[-V_D(m_{0}^2h^2/2+\lam h^4/4)]}{\int\di^\mathcal{N}h\,\exp[-V_D(m_{0}^2h^2/2+\lam h^4/4)]}\n\\
=&~\FR{2}{\sqrt{V_D\lam}}\FR{{}_1\wt F_1\big(\frac{\mathcal{N}+2}{4};\fr{1}{2};z^2\big)-z{\;}_1\wt F_1\big(\frac{\mathcal{N}+4}{4};\fr{3}{2};z^2\big)}{{}_1\wt F_1\big(\frac{\mathcal{N}}{4};\fr{1}{2};z^2\big)-z{\;}_1\wt F_1\big(\frac{\mathcal{N}+2}{4};\fr{3}{2};z^2\big)},
\end{align}
in which $z\equiv \frac{1}{2}m_{0}^2\sqrt{V_D/\lam}$,  ${}_1\wt F_1(a;b;z)\equiv \frac{\Gamma(a)}{\Gamma(b)}{}_1F_1(a;b;z)$, and ${}_1F_1(a;b;z)$ is hypergeometric function of type-$(1,1)$. We can extract the effective mass $m_{\text{eff}}^2$ from this 2-point function of zero modes by expressing it in terms of spherical harmonics,
\bge
\label{hihj}
  \la h_i h_j \ra=\de_{ij}\FR{H^{D-2}Y_{\vec 0}^2}{(m_{\text{eff}}/H)^2}=\de_{ij}\FR{1}{V_Dm_{\text{eff}}^2}.
\ede
Then it is clear that $m_{\text{eff}}^2=\mathcal{N}\big(V_D\la h^2\ra\big)^{-1}$.

The above calculation shows how scalar fields acquire $O(\mathcal{N})$-symmetric mass correction. In the following section, we shall also deal with symmetry broken gauge theories, where the would-be Goldstone components of scale fields are transferred to the longitudinal polarizations of the gauge field. In this case, it is mostly convenient to go to unitary gauge. Here let us illustrate this case with a specific symmetry breaking pattern $SU(2)\to U(1)$ and consider an $SU(2)$ doublet scalar $\Phi$. We can parameterize $\Phi = \frac{1}{\sqrt{2}}\wh{h} e^{\ii \wh{\pi}^i\si^i}$ with $\wh{\pi}^i$ the three would-be Goldstone components and $\si^i$ the standard Pauli matrices. We have used hatted variables to denote fields in unitary gauge. Then there is only one real physical component $\wh{h}$ in $\Phi$, and the 2-point correlator of its zero mode is evaluated as follows,
\begin{align}
\label{h2hat}
  \la \wh h^2\ra=&~\FR{\int\di^3\wh \pi\int\di\wh h\,\wh h^3\times \wh h^2\exp[-V_D(m_{0}^2\wh h^2/2+\lam \wh h^4/4)]}{\int\di^3\wh \pi\int\di \wh h\,\wh h^3\,\exp[-V_D(m_{0}^2\wh h^2/2+\lam \wh h^4/4)]},\end{align}
where we see that $\wh \pi^i$'s disappear from the action and the path integral of their zero modes factors out. As a result, the above correlator is identical to $\la h^2\ra$ in (\ref{h2vev}), with $\mathcal{N}=4$. In the similar way, the effective mass of $\wh h$ is also given by $m_{\text{eff}}^2=\mathcal{N}(V_D\la \wh h^2\ra)^{-1}$. In Sec.~\ref{Sec_mass} we shall use this unitary gauge result to determine the Higgs mass during inflation where we shall also explain the classical origin of a nonzero $m_0$ even in the absence of quadratic term of Higgs potential.

Finally we note that the $\la h_i h_i\ra$ correlator (no summation over $i$) in (\ref{hihj}) and $\la\wh h^2\ra$ correlator in (\ref{h2hat}) differ by a factor of $\mathcal{N}$. In the case of $SU(2)$, this would be a factor of 4. If we consider an $U(1)$ gauge theory with one complex scalar instead, the difference would be a factor of 2. We emphasis this seemingly trivial difference because this fact will be crucial in Sec.\;\ref{SubSecLCVBM} when we show that the mass correction is independent of gauge choice.

\subsection{Loop Correction to Fermion Mass}

Next we consider the 1-loop mass correction to a Dirac spinor. As mentioned above, the only possible mass correction come from the scalar loop. Thus let's consider the following action with a real scalar and a Dirac fermion interacting through Yukawa term,
\bge
  S=\int\di^4x\det(e_\nu^n)\,\bigg[\FR{\ii}{2}\Big(\ob{\psi}\gamma^\mu\nabla_\mu\psi-\ob{\psi}\overleftarrow{\nabla}_\mu\gamma^\mu\psi\Big)-\FR{1}{2}(\pd_\mu\phi)^2-\FR{1}{2}m^2\phi^2-y\phi\ob\psi\psi\bigg],
\ede
where $e_\nu^n$ is the vierbein, and $\nabla_\mu$ is the standard covariant derivative containing spin connection term. For simplicity we only consider the case of massless fermion, which is enough for our purpose since SM fermions are all massless so long as the Higgs field does not acquire nonzero background value.

We are interested in a possible 1-loop contribution to the Dirac mass term $\ob{\psi}\psi$. To this end we need to evaluate the following loop integral,
\bge
\Delta S=y^2\int\di\Omega\di\Omega'\,\ob{\psi}(x)G_F(x,x')G_{\phi}(x',x)\psi(x').
\ede
We do not bother to spell out explicitly the renormalization scale $\mu_R$ dependence when $D\neq 4$ because it does not play any role in our calculation. As in the previous case, to find the loop correction to the Dirac mass term, it is enough to set the two external fields $\ob\psi(x)$ and $\psi(x')$ to be constant. Then it is straightforward to show that this integral actually vanishes. In fact,
\begin{align}
\Delta S=&~y^2H^{2D-2}\sum_{\vec L,\vec M,s}\int\di\Omega\di\Omega'\,\ob{\psi}(x)\sla X\n\\
&~\times\bigg[\FR{1}{\lam_L^+}Y_{\vec L s}^+(x)Y_{\vec L s}^{+\dag}(x')+\FR{1}{\lam_L^-}Y_{\vec L s}^-(x)Y_{\vec L s}^{-\dag}(x')\bigg]\FR{1}{\lam_M}Y_{\vec M}^*(x)Y_{\vec M}(x')\psi(x')\n\\
=&~y^2H^{D-2}\sum_{\vec L,s}\int\di\Omega\,\ob{\psi}(x)\sla X\bigg[\FR{1}{\lam_L^+\lam_L^{}}Y_{\vec L s}^{+}(x)\psi_s^\dag Y_{\vec L}(x)+\FR{1}{\lam_L^-\lam_L^{}}Y_{\vec L s}^{-}(x)\psi_s^\dag Y_{\vec L}(x)\bigg]\psi(x)\n\\
=&~\FR{-y^2H^{D-2}}{\mu^2-1/4}\sum_{\vec L,s}\int\di\Omega\,\ob{\psi}(x)\sla X\n\\
&~\times\bigg[\Big(\FR{1}{\lam_L^+}-\FR{1+\lam_L^+}{\lam_L}\Big)Y_{\vec L s}^{+}(x)\psi_s^\dag Y_{\vec L}(x)+\Big(\FR{1}{\lam_L^-}-\FR{1+\lam_L^-}{\lam_L}\Big)Y_{\vec L s}^{-}(x)\psi_s^\dag Y_{\vec L}(x)\bigg]\psi(x)\n\\
=&~\FR{-y^2}{\mu^2-1/4}\int\di\Omega\,\ob{\psi}(x)\Big[H^{-2}G_F(x,x)-\Big(\sla X+H^{-2}\sla\nabla_x)G_\phi(x,x')\Big)_{x'=x}\Big]\psi(x)\n\\
=&~0,
\end{align}
where $\lam_L$ is given in (\ref{ScaProSH}), $\lam^\pm$ is given in (\ref{SpSH}), and $\psi_s$ is a basis for Dirac spinors. In above derivation we have used the definition of spin-weighted spherical harmonics (\ref{SpSH}), the orthonormal condition of spin-weighted spherical harmonics (\ref{SpSHON}), and the relation (\ref{SpSHDecomp}). The final expression must vanish on symmetry ground. To see this point more explicitly, we note that each term in the second-to-last line of above expression contains either $\ga^\mu X_\mu$ or $\ga^\mu \pd_\mu$, which can be further written as $\ga^m e_m^\mu(x)X_\mu$ or $\ga^m e_m^\mu(x)\pd_\mu$. In these expressions, $\ga^m$ is coordinate independent and thus can be taken out of the integral. The rest factor is then a dS-invariant quantity with single vector index and thus must be zero.  Therefore, we see that the 1-loop correction to fermion's mass vanishes if the fermion itself is classically massless. This confirms the results found in \cite{Chen:2016nrs}, and also agrees with the flat space result.

\subsection{Loop Correction to Vector Boson Mass}
\label{SubSecLCVBM}

Finally, we consider the correction to Abelian gauge boson. Here we consider the action of a complex real scalar charged under a $U(1)$ gauge symmetry,
\begin{align}
  S=&-\int\di^4x\sqrt{-g}\,\Big[\FR{1}{4}F_{\mu\nu}F^{\mu\nu}+|\D_\mu\Phi|^2+m^2\Phi^\dag\Phi\Big]\n\\
  \supset&-\int\di^4x\,\sqrt{-g}\bigg[\FR{1}{2}e^2 A^2(\pi^2+\phi^2)+e A^\mu(\phi\pd_\mu\pi-\pi\pd_\mu\phi)\bigg],
\end{align}
where we have parameterized the complex scalar field $\Phi$ in terms its two real components as $\Phi=\frac{1}{\sqrt 2}(\phi+\ii\pi)$. We do not need to specify the gauge fixing condition explicitly at this place because we won't use gauge boson's propagator. However, for the sake of rigorousness, we can just choose Lorentz gauge $\nabla_\mu A^\mu=0$ and follow the standard Faddeev-Popov quantization, so that the two real components of scalar field have the mass as indicated in the Lagrangian. It's also worth mentioning that the Faddeev-Popov ghost does not contribute to 1-loop correction of Abelian gauge boson's two-point function, because the ghost field, which has the same charge as the gauge transformation parameter, is neutral in Abelian theory. This is no longer the case in non-Abelian gauge theory where the ghost is charged under gauge group and interacts with gauge field. This would make the calculation for non-Abelian theory more complicated. Fortunately, at the end of this section, we shall show that the same result of mass correction can also be obtained in unitary gauge, where only physical degrees of freedom appear.

In the second line of above expression we show explicitly the two types of interactions that contribute to 1-loop correction of photon's 2-point function. They contribute to the following two types of diagrams, respectively,
\bge
\label{vec2pt}
\parbox{0.6\textwidth}{\vspace{-1mm}\includegraphics[width=0.55\textwidth]{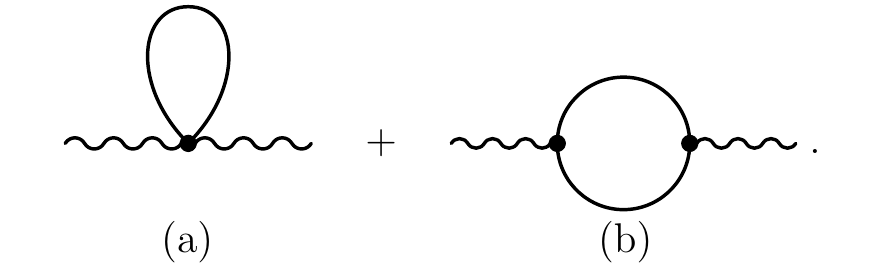}}
\ede
The Diagram (\ref{vec2pt}a) is similar to the case of $\phi^4$ loop and is easy to calculate. As in \cite{Chen:2016nrs}, this diagram contributes nonzero mass to the gauge boson, i.e., a term proportional to $A_\mu A^\mu$,
\begin{align}
\label{Vec2ptA}
\text{Diag.\,\ref{vec2pt}a}=&~e^2\mu_R^{4-D}\Big[G_\phi(x,x)+G_\pi(x,x)\Big]\n\\
 =&~\FR{e^2\mu_R^{4-D}H^{D-2}}{2(4\pi)^{D/2-1}}\FR{\Gamma(\mu+\fr{d}{2})\big[\tan(\fr{\pi d}{2})-\cot(\pi\mu-\fr{\pi d}{2})\big]}{\Gamma(\mu-\fr{d}{2}+1)\Gamma(\frac{1+d}{2})}.
\end{align}
Expanding the above expression around $\ep\equiv3-d=0$, we get,
\begin{align}
\label{Vec2ptDiagA}
  \text{Diag.\,\ref{vec2pt}a}=&~\FR{e^2(2H^2-m^2)}{8\pi^2}\FR{2}{\ep}-\FR{e^2}{8\pi^2}\bigg[m^2+(1+2\mu)H^2+(2H^2-m^2)\n\\
  &~\times\Big(\ga_E+2\psi(\mu-\FR{1}{2})+\log\FR{H^2}{\mu_R^2}-\log4\pi-\pi\tan(\pi\mu)\Big)\bigg]+\order{\ep}.
\end{align}
Here it is more convenient to use Minimal Subtraction (MS) rather than $\ob{\text{MS}}$ scheme. Under MS scheme, the mass correction to the photon from $A^2\phi^2$ interaction is simply the $\order{\ep^0}$ terms of the above equation,
\begin{align}
\label{deMADiaga}
  \de M_A^2(\text{Diag.\,\ref{vec2pt}a})=&-\FR{e^2}{8\pi^2}\bigg[m^2+(1+2\mu)H^2+(2H^2-m^2)\n\\
  &~\times\Big(\ga_E+2\psi(\mu-\FR{1}{2})+\log\FR{H^2}{\mu_R^2}-\log4\pi-\pi\tan(\pi\mu)\Big)\bigg].
\end{align}
Then in the small scalar mass limit $m/H\ll 1$, the mass correction (\ref{deMADiaga}) becomes,
\bge
\label{deMAexpand}
  \de M_A^2(\text{Diag.\,\ref{vec2pt}a})=\FR{3e^2H^4}{4\pi^2m^2}+\order{m^0},
\ede
which reduces to the result found in \cite{Chen:2016nrs}\footnote{In \cite{Chen:2016nrs} the mass correction to gauge boson $\de M_A$ can never be larger than $H/2$, due to the limitation of dynamical renormalization group resummation. Here we see that (\ref{deMAexpand}) holds even when $m\ll H$, and thus the mass correction $\de M_A$ can probably exceed $H/2$. However, one caveat is that the loop expansion may break down for very small $m/H$, because the expansion parameter here is actually $(eH/4\pi m)^2$ rather than the typical $(e/4\pi)^2$. Therefore, even the 1-loop calculation here indicates that $\de M_A$ can be large, it should be noted that the 1-loop result alone is no longer a good approximation when $m\ll eH$.}. But in order to fully justify this result, it remains to be seen that Diag.\,\ref{vec2pt}b's contribution can be ignored in the $m\ll H$ limit. Diag.\,\ref{vec2pt}b is more difficult to compute than \ref{vec2pt}a. But in the limit of vanishing external momentum and using some tricks, we can get analytic expression for it. To show this, we firstly write down the expression of Diag.\,\ref{vec2pt}b in the limit of vanishing external momentum,
\begin{align}
\label{Vec2ptB}
2e^2\mu_R^{4-D}A^\mu(x)A^{\mu'}(x') \int\di\Omega\di\Omega'\,\Big[G_{\phi}(x,x')\pd_\mu\pd_{\mu'} G_{\phi}(x,x')-\pd_\mu G_{\phi}(x,x')\pd_{\mu'} G_{\phi}(x,x')\Big],
\end{align}
where we have treated $A^\mu(x)$ and $A^{\mu'}(x')$ as constants and pulled them out of integral\footnote{Here we encounter the topological obstacle from the hairy ball theorem which states that there is no nonvanishing smooth vector field over $S^4$. But we can refine our choice of field configuration by requiring that $A^\mu$ is a constant vector field over $S^4$ except at the north pole and south pole, at which we require $A^\mu$ to be zero. Then one can readily check that all following calculation is not affected by the removal of two polar points.}. It is still quite difficult to carry out the above integral by brute force. Fortunately, we are able to simplify the calculation by using symmetry arguments.
On symmetry ground, the contribution to a local operator from above integral must be proportional to $g_{\mu\mu'}$. Therefore, to get the result of above integration, it is enough to calculate one component, which we choose to be the azimuthal direction $g_{\varphi\varphi}$. This is a great simplification because the spherical harmonics $Y_{\vec L}(x)$ are eigenfunctions of $\pd_\varphi$, that is, $\pd_\varphi Y_{\vec L}(x)=\ii L_1 Y_{\vec L}(x)$ where $L_1$ is the ``last'' component of the vector $\vec L=(L,L_{D-1},L_{D-2},\cdots,L_1)$. Then using (\ref{ScaProSH}), we further have $\pd_\varphi G(x,x')=-\pd_{\varphi'}G(x,x')$, where $\varphi$ and $\varphi'$ denotes the azimuthal directions at point $x$ and $x'$, respectively. Therefore, the $\varphi\varphi$-component of above integral (with $A^\varphi A^{\varphi'}$ suppressed) can be rewritten in the following way using (\ref{ScaProSH}),
\begin{align}
   &-2e^2\mu_R^{4-D}H^{2D-2}\int\di\Omega\di\Omega'\sum_{\vec L,\vec M}\FR{1}{\lam_L\lam_M}\n\\
   &~~\times\Big[Y_{\vec L}^{}(x)Y_{\vec L}^*(x')Y_{\vec M}^{}(x')\pd_\varphi^x\pd_\varphi^x Y_{\vec M}^*(x)-\big(\pd_\varphi^xY_{\vec L}^{}(x)\big)Y_{\vec L}^*(x')Y_{\vec M}^{}(x')\pd_\varphi^x Y_{\vec M}^*(x)\Big]\n\\
  =&-4e^2\mu_R^{4-D}H^{D-2}\int\di\Omega\sum_{\vec L}\FR{1}{\lam_L^2} Y_{\vec L}^{}(x)\pd_\varphi^x\pd_\varphi^x Y_{\vec L}^*(x)\n\\
  =&~4e^2\mu_R^{4-D}H^{D-2}\int\di\Omega\sum_{\vec L}\FR{L_1^2}{\lam_L^2} Y_{\vec L}^{}(x) Y_{\vec L}^*(x)\n\\
  =&-4e^2\mu_R^{4-D} \FR{\pd}{\pd m^2}\sum_{\vec L}\FR{L_1^2}{\lam_L}.
\end{align}
Then we compare this with the tree level diagram from $g_{\varphi\varphi}A^\varphi A^\varphi$ component of the mass term with $A^\varphi$ constant,
\bge
  -2\times\FR{1}{2}\int\di\Omega\,M_A^2g_{\varphi\varphi}(x)A^\varphi A^\varphi=-\FR{2M_A^2}{D+1}V_D A^\varphi A^\varphi,
\ede
It is straightforward to see that the mass correction from Diagram \ref{vec2pt}b is,
\begin{align}
\label{Vec2ptBsum}
  &\de M_A^2(\text{Diag.\,\ref{vec2pt}b})
  =2e^2\mu_R^{4-D}(D+1)V_{D}^{-1}\FR{\pd}{\pd m^2}\sum_{\vec L}\FR{L_1^2}{\lam_L} \n\\
  &=-\FR{4e^2H^{D}\mu_R^{4-D}\Gamma(2+d/2)}{\pi^{d/2+1}\Gamma(3+d)\cos(\pi d/2)}\FR{\pd}{\pd m^2}\Big[\cos(\pi\mu)\Gamma(1+\FR{d}{2}+\mu)\Gamma(1+\FR{d}{2}-\mu)\Big],
\end{align}
where the summation over $\vec L$ can be carried out in closed form, similar to what we have done in the toy example of Sec.\;\ref{sec_toy},
\begin{align}
\label{Lsum2}
\sum_{\vec L}\FR{L_1^2}{\lam_L}=&~\sum_{L_D=0}^\infty\sum_{L_{D-1}=0}^{L_D}\cdots\sum_{L_2=0}^{L_3}\sum_{L_1=-L_2}^{L_2}\FR{L_1^2}{\lam_{L_D}}\n\\
=&\sum_{L_D=0}^\infty\FR{1}{\lam_D}\FR{2(2L_D+d)\Gamma(L_D+d+1)}{\Gamma(d+3)\Gamma(L_D)}\n\\
=&-\FR{2\cos(\pi\mu)\Gamma(1+d/2+\mu)\Gamma(1+d/2-\mu)}{\cos(\pi d/2)\Gamma(d+3)}.
\end{align}

Now we expand the above result around $\ep\equiv 3-d=0$,
\begin{align}
\label{Vec2ptDiagB}
  \text{Diag.\,\ref{vec2pt}b}=&~\FR{e^2(-H^2+m^2)}{8\pi^2}\FR{2}{\ep}\n\\
  &+\FR{e^2}{16\pi^2}\FR{\pd}{\pd m^2}\bigg\{(2H^2-m^2)m^2\Big[-\FR{3}{2}+\ga_E-\log(4\pi)+\log\FR{H^2}{\mu_R^2}\n\\
  &-(\FR{3}{2}-\mu)\psi(\FR{5}{2}-\mu)+(\FR{3}{2}+\mu)\psi(\FR{5}{2}+\mu)-\FR{3}{2}\pi\tan(\pi\mu)\Big]\bigg\}+\order{\ep}.
\end{align}
Then, under MS scheme, we get the final expression for the mass correction from Diag.\,\ref{vec2pt}b to be,
\begin{align}
\label{deMB}
&\de M_A^2(\text{Diag.\,\ref{vec2pt}b})\n\\
=&~\FR{e^2}{16\pi^2}\FR{\pd}{\pd m^2}\bigg\{(2H^2-m^2)m^2\Big[-\FR{3}{2}+\ga_E-\log(4\pi)+\log\FR{H^2}{\mu_R^2}\n\\
  &-(\FR{3}{2}-\mu)\psi(\FR{5}{2}-\mu)+(\FR{3}{2}+\mu)\psi(\FR{5}{2}+\mu)-\FR{3}{2}\pi\tan(\pi\mu)\Big]\bigg\}.
\end{align}
To justify the $m\ll H$ limit expression in (\ref{deMAexpand}), we need to show that (\ref{deMB}) is subdominant in this limit. Indeed, after expanding in $m/H$, we get
\bge
\label{dMA2Exp}
  \de M_A^2(\text{Diag.\,\ref{vec2pt}b})=\FR{e^2H^2}{8\pi^2}\Big(\FR{10}{3}-2\ga_E+\log4\pi+\log\FR{H^2}{\mu_R^2}\Big)+\order{m^2/H^2}.
\ede
Therefore, the mass contribution from this diagram is finite as $m\to 0$, which confirms the $m\ll H$ limit expression (\ref{deMAexpand}), and thus the result found in \cite{Chen:2016nrs}.

To illustrate the contributions from different diagrams, we plot the mass corrections from Diagrams (\ref{vec2pt}a) and (\ref{vec2pt}b), as well as their sum, as functions of loop scalar mass $m$, in Fig.\;\ref{FigPhoMass}, from which one can see clearly that Diagram (\ref{vec2pt}a) is divergent when $m\to 0$ while Diagram (\ref{vec2pt}b) remains finite.

\begin{figure}[tbph]
\centering
\includegraphics[width=0.6\textwidth]{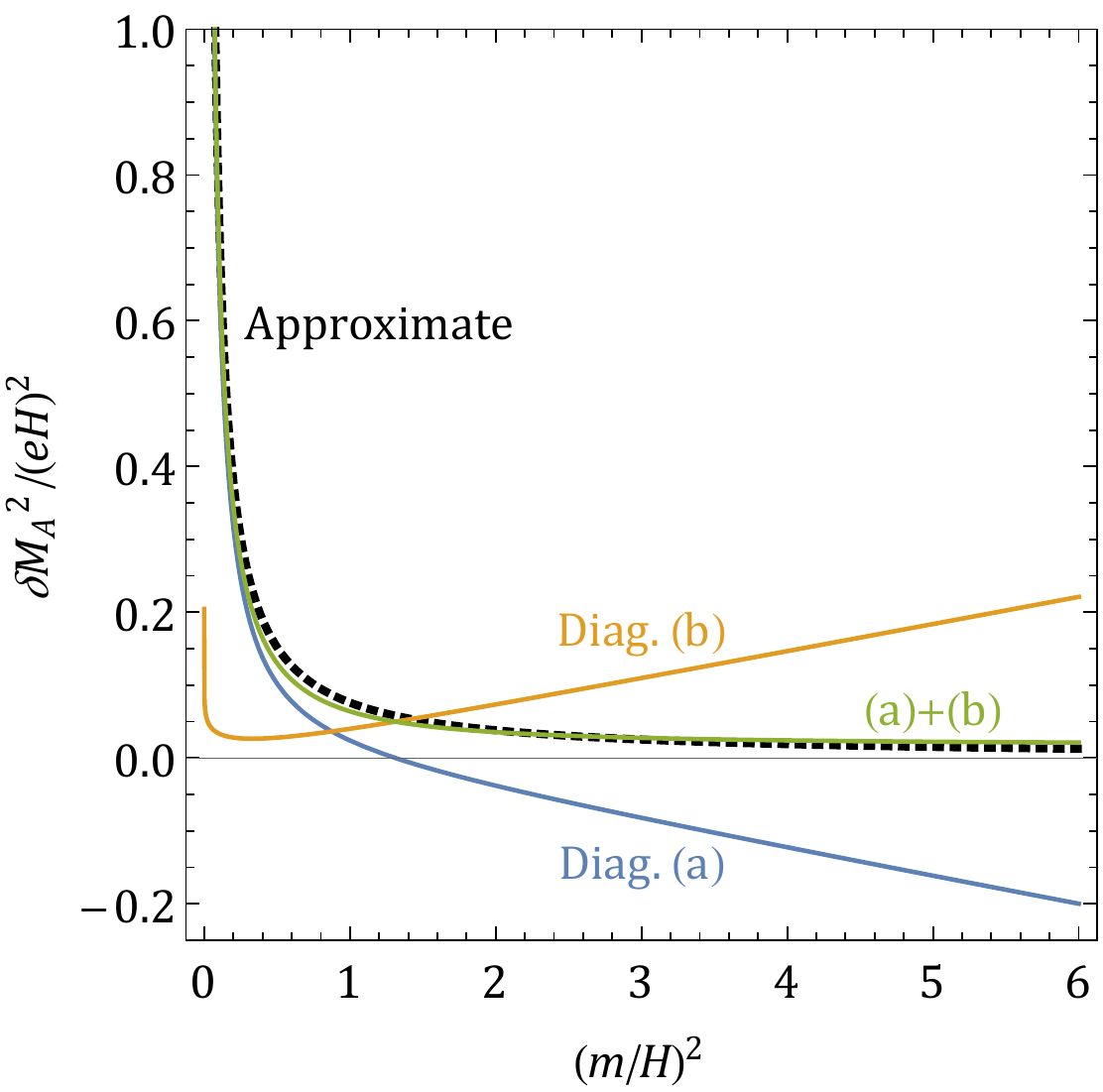}
\caption{The 1-loop mass corrections to the gauge boson $\de M_A^2$ as functions of the mass of the loop scalar field $m^2$. The blue, orange, and green curves correspond to the mass corrections from diagram (\ref{vec2pt}a) [given by Eq.~(\ref{deMADiaga})], diagram (\ref{vec2pt}b) [given by Eq.~(\ref{deMB})], and the sum of the two, respectively. The dashed line is the approximation Eq.~(\ref{deMAexpand}).}
\label{FigPhoMass}
\end{figure}

The interesting point to note is that although (\ref{deMAexpand}) is only the leading order term of mass correction in the $m\ll H$ limit, it is nevertheless a very good approximation even when $m$ is comparable with $H$, as is clear from Fig.\,\ref{FigPhoMass}. Below we shall also show that (\ref{deMAexpand}) actually corresponds to the contribution of zero mode. When $m>H$, there are some discrepancy between the approximate result (\ref{deMAexpand}) and the full result (\ref{deMADiaga})$+$(\ref{deMB}). But it should also be noted that the mass correction for $m>H$ is also subject to uncertainties from the choice of renormalization scale. Therefore, unless one demands very high precision, we can just use (\ref{deMAexpand}) without considering much more complicated full expression (\ref{deMADiaga}) and (\ref{deMB}).

In Fig.\;\ref{FigPhoMass} one can also observe that both Diagrams (\ref{vec2pt}a) and (\ref{vec2pt}b) increase as $m/H$ goes large. However, there are some cancellation between the two diagrams. This is actually an important consistency check of our result, which we would like to spell out more explicitly here. The consistency with flat space limit $H\to 0$ requires that both the divergence part (proportional to $1/\ep$) and the finite part (proportional to $\ep^0$) cancel out between the two diagrams, because the gauge field must remain massless as required by gauge invariance\footnote{One caveat to this argument is that spontaneous gauge symmetry breaking may happen as in the case of Coleman-Weinberg model. But we can exclude this case by taking the mass of the scalar field large enough.}. Now it is obvious from (\ref{Vec2ptDiagA}) and (\ref{Vec2ptDiagB}) that the terms proportional to $1/\ep$ in both diagrams cancel each other when $H=0$. But it is less obvious that the finite part (proportional to $\ep^0$) in (\ref{Vec2ptDiagA}) and (\ref{Vec2ptDiagB}) also cancel each other in $H\to 0$ limit. To see this is indeed the case, we expand both expressions in $m\gg H$ limit, using $\mu\simeq \ii m/H$ and the asymptotic behavior of digamma function $\psi(z)\sim \log(z)+\order{z^{-1}}$, and we find,
\begin{align}
&\de M_A^2(\text{Diag.\,\ref{vec2pt}a})\n\\
=&~\FR{e^2m^2}{8\pi^2}\Big(-1+\ga_E-\log 4\pi+\log\FR{m^2}{\mu_R^2}\Big)+\order{H}\n\\
=&-\de M_A^2(\text{Diag.\,\ref{vec2pt}b}).
\end{align}
Therefore, we see that the mass correction in flat space limit $H\to 0$ does vanish due to the cancellation of two diagrams.

In this subsection, we have performed the loop calculation in a diagram-wise manner. On the other hand, we recall that the IR mass generation in de Sitter is usually attributed to the divergence of zero mode of the scalar field. Therefore, it is also illuminating to recast our calculation of gauge boson's mass correction in mode-wise manner. That is, we want to rewrite the 1-loop mass correction to gauge boson as a summation over modes of the loop scalar. This can be conveniently done by rewriting (\ref{Vec2ptA}) and (\ref{Vec2ptBsum}) in the following form with the help of (\ref{Lsum1}) and (\ref{Lsum2}),
\begin{align}
\label{dMAmodeA}
&\de M_A^2(\text{Diag.}\;\ref{vec2pt}a)=\FR{2e^2\mu_R^{4-D}}{V_DH^2}\sum_L \Delta_L(a),
&&  \Delta_L(a)=\FR{1}{\lam_L}\FR{(2L+d)\Gamma(L+d)}{\Gamma(d+1)\Gamma(L+1)},\\
\label{dMAmodeB}
&\de M_A^2(\text{Diag.}\;\ref{vec2pt}b)=\FR{2e^2\mu_R^{4-D}}{V_DH^2}\sum_L \Delta_L(b),
&& \Delta_L(b)=\FR{1}{\lam_L^2}\FR{2(2L+d)\Gamma(L+d+1)}{\Gamma(d+2)\Gamma(L)},
\end{align}
where $\lam_L=L(L+d)+(m/H)^2$.

Now we make several remarks about these expressions. Firstly, when $m\ll H$, the zero mode ($L=0$) contributes dominantly to the mass correction in Diagram \ref{vec2pt}a and the contribution scales as $1/m^2$, while the zero-mode contribution in Diagram \ref{vec2pt}b is constantly zero, due to $\Gamma(L)$ factor in the denominator of $\Delta_L(b)$. This must be so because the zero mode represents the constant component of the loop fields, and thus it must vanish when there are spacetime derivatives acting on loop lines as in (\ref{Vec2ptB}). In particular, one can immediately recognize that the approximate expression (\ref{deMAexpand}) can be got by keeping only the $L=0$ terms in (\ref{dMAmodeA}).

Secondly, the nonzero-mode contributions from two diagrams almost (though not exactly) cancel each other, giving negligible contribution to the mass correction.
To make this point more clear, we plot $\Delta_L(a)$, $\Delta_L(b)$, as well as their sum, for first several modes, in Fig.\;\ref{FigModeSum}, choosing the mass $m=\frac{1}{5}H$. It should not be a surprise that the mass contribution in each nonzero mode is positive for Diagram \ref{vec2pt}a (negative for \ref{vec2pt}b), while from Fig.\;\ref{FigPhoMass} it is clear that the summation over all modes sometimes gives negative contribution for \ref{vec2pt}a (and positive contribution for \ref{vec2pt}a). This is because the results presented in Fig.\;\ref{FigPhoMass} are already dimensional-regularized, and we know that a positive but divergent series can indeed sum to a negative value after regularization.

\begin{figure}[tbph]
\centering
\includegraphics[width=0.65\textwidth]{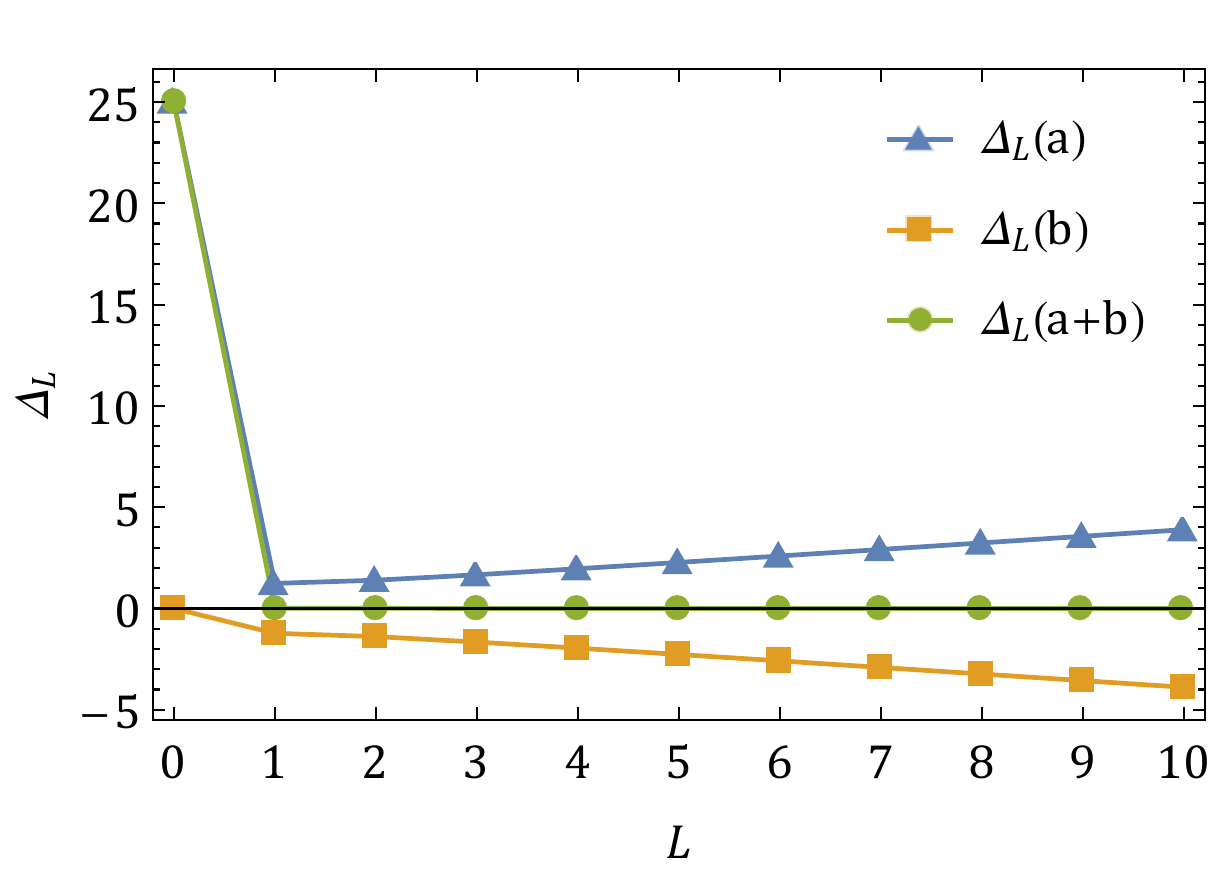}
\caption{The 1-loop correction to photon's mass from a scalar with mass $m=\frac{1}{5}H$, as a function of angular quantum number $L$ of loop modes. The function $\Delta_L$ is defined as in (\ref{dMAmodeA}) and (\ref{dMAmodeB}).}
\label{FigModeSum}
\end{figure}

Finally but not least importantly, we note that the mass correction to photon can be attributed to the scalar zero mode contribution. In this process, the scalar field itself does not pick up a nonzero VEV. On the other hand, the zero-mode-squared does pick up a nonzero VEV, see (\ref{h2vev}). As a result, the gauge invariance of the kinetic term of the scalar field $|\D_\mu\mb \phi|^2$ is broken, due to the nonvanishing 2-point function of scalar zero mode, and it is for this reason that the gauge boson becomes massive. Therefore, the mass generation mechanism here is quite different from the standard Higgs mechanism, although we still expect that one of scalar degree of freedom is converted to the longitudinal polarization of the gauge boson.

The disappearance of would-be Goldstone component in the scalar field inspires us to revisit the above calculation in unitary gauge, where we parameterize the complex scalar field as $\Phi=\wh\phi e^{\ii\wh\pi}$. It is well known that unitary gauge is not suitable for loop computation as the structure of UV divergence in this gauge is obscure. If we view the unitary gauge as the $\xi\to\infty$ limit of $R_\xi$ gauge, then this problem can sometimes be viewed as the noncommutativity of taking loop integral and taking $\xi\to\infty$ limit. This is manifest in our result, because if we turn off $\pi$ field directly, then Diagram (\ref{vec2pt}b) will disappear, and the crucial cancellation of UV divergence between Diagram (\ref{vec2pt}a) and (\ref{vec2pt}b) in the flat space limit no longer holds.

However, so far as we are concerned with the leading mass correction from zero modes, we should be able to recover (\ref{deMAexpand}) even in unitary gauge, because the zero mode has no UV divergence so unitary gauge should work in principle. This is indeed true: if we only focus on zero modes in unitary gauge, then $\wh\pi$ field disappears completely. Then we will have Diagram (\ref{vec2pt}a) only, with only $\wh\phi$ field running in the loop. So apparently the mass correction should be half of (\ref{deMAexpand}). However, as emphasized at the end of Sec.\;\ref{SecZM}, the zero mode 2-point correlator $\la\wh\phi^2\ra$ in the unitary gauge is twice of the correlator $\la \phi^2\ra$ in Lorentz gauge we used above, so the two factors cancel out, and we conclude that the zero-mode contribution to gauge field mass is the same in both Lorentz gauge and unitary gauge. We shall make use of this agreement in next section to compute SM gauge bosons' mass with unitary gauge, because the computation in Lorentz gauge is more involved for non-Abelian gauge theory due to Faddeev-Popov ghost.

At the end of this subsection, we mention a possible concern regarding our method of calculating the mass correction. In this section, we have turned off the external momentum, and focused on the momentum independent part of the loop correction. In particular, we interpret the loop correction proportional to $A_\mu A^\mu$ as the correction to photon's mass. However, in dS, the kinetic term of photon also contributes a term proportional to $A_\mu A^\mu$, due to the nonzero constant background curvature. In fact, after integration by parts, one can find the following quadratic terms from the kinetic term of photon,
\bge
  \FR{1}{4}\sqrt{g}F_{\mu\nu}F^{\mu\nu}=\FR{1}{2}\sqrt{g}\Big[A_\mu\big(-\nabla^2+3H^2\big)A^\mu-A^\mu\nabla_\mu\nabla_\nu A^\nu\Big],
\ede
where $\nabla^2\equiv g^{\mu\nu}\nabla_\mu\nabla_\nu$, and there is an apparent mass-like term with squared ``mass'' $3H^2$. Then one may wonder whether our result is actually a wave function renormalization (i.e. a correction to kinetic term $F_{\mu\nu}^2$) rather than mass correction, and that it appears to be a mass term proportional to $A_\mu A^\mu$ just because we have turn off the external momentum.

In general, this question can only be answered by keeping the external momentum finite, which however would make the computation much more complicated. Fortunately, for our result (\ref{deMAexpand}) which comes predominantly from zero modes of loop scalar as elaborated above, there is a simple way to see that it is a genuine mass correction rather than wave function renormalization. In fact, if (\ref{deMAexpand}) is a wave function renormalization, then there must be a corresponding correction to $A_\mu\nabla^2 A^\mu$. Such a contribution can arise only from Diagram (\ref{vec2pt}b). But we have seen that zero modes contribute nothing to this diagram, and this remains correct even when we turn on the external momentum. Therefore, we conclude that the expression (\ref{deMAexpand}) is a genuine correction to photon's mass rather than a wave function renormalization.

\section{SM Spectrum in Non-Higgs Inflation}
\label{Sec_mass}

The SM mass spectrum during inflation can be quite different from the the familiar SM spectrum, i.e. the spectrum in the electroweak broken phase. There are several new contributions to the masses of SM particles which we are going to clarify in this section. In this paper we shall focus on the single field slow-roll models for simplicity, assuming that the inflaton has effective couplings with the SM sector parameterized by some unknown functions. The analysis can be generalized to more general effective field theories of inflation models.

Before entering the details, here we present an overview of new ingredients that would contribute the SM spectrum during inflation. The first and foremost ingredient is the background value of Higgs field. In electroweak broken vacuum, the Higgs VEV $v_h\simeq 246$GeV is nonzero due to the negative quadratic term in the Higgs potential. In typical inflation models, however, this scale is too low and thus can be safely neglected in most cases. Then, given a positive quartic potential for Higgs field, one may expect that the Higgs VEV is constantly zero during inflation, so long as the inflation scale is high enough. However, we have two exceptions. Firstly, in Higgs inflation such as the one we will study in Sec.~\ref{Sec_Higgs}, the Higgs field is identical to the inflaton field, and it picks up a large VEV during inflation, which can be of $\order{\Mp}$. Secondly, even in non-Higgs inflation scenarios, we may also have a nonzero Higgs VEV if there is spontaneously symmetry breaking, which we will comment at the end of Subsection~\ref{sec:fvbm}. In this and next sections, we shall mainly focus on the non-Higgs inflation scenarios with zero Higgs VEV.
For non-Higgs inflation, we only need to consider the soft fluctuations of Higgs field, which have been consistently taken account of in our treatment of zero modes of Higgs fields in Sec.\;\ref{SecZM} in form of quantum corrections, so we can conveniently include these quantum corrections after working out the classical mass spectrum.

The second ingredient is the nonminimal coupling between Higgs field and the Ricci scalar,
\bge
  S\supset-\int\di^4x\sqrt{-g}\xi R\mb H^\dag\mb H,
\ede
which is the unique dim-4 operator in the effective theory of SM + general relativity, and represents the leading term of a whole series of effective operators between SM fields and gravitational fields. This term is particularly important for determining the Higgs mass because it introduces a tree level mass $\xi R=12\xi H^2$ to the Higgs. Since other SM fields may receive mass from Higgs loop, which depends on Higgs mass in an important way, this operator is also potentially important for determining the masses of other SM particles. In principle, one can also consider higher dimensional operators between SM and gravity, but they are expected to be suppressed by a very high scale, e.g. Planck scale if we assume general relativity for the gravitational sector. Therefore, we shall ignore them in this paper.

The third ingredient is the coupling between the inflaton field $\phi$ and the SM fields. These interactions can be quite arbitrary, and one can in principle specify them in each given inflation model. Here we choose not to specify the inflation model, but to proceed with a very broad class of inflaton-SM interactions. If we assume that the inflaton field is SM singlet, then we can write down very general couplings between the inflaton and SM fields in the following way,
\bge
\label{InfOSM}
  S\supset -\int\di^4x\sqrt{-g}\sum_\al f_\al(X,\phi)\O_\al[\Phi_\text{SM}],
\ede
where $f_{\al}(X,\phi)$ are arbitrary functions of $X\equiv(\pd_\mu\phi)^2$ and $\phi$, which we assume to be well behaved in the sense that a sensible low energy limit should be recovered, and $\OSM$ is any singlet operator constructed from SM fields, collectively denoted by $\Phi_\text{SM}$.

The interaction (\ref{InfOSM}) may introduce a number of uncertainties in our study though. Taking a term $f_H(X,\phi)\mb H^\dag \mb H$ from (\ref{InfOSM}) for example, the above interaction would introduce additional squared mass $\de M_H^2=f_H(X_0,\phi_0)$ with $X_0\equiv-\dot\phi_0^2$. As another example, the coupling $f_{DH}(X,\phi)|\D_\mu\mb H|^2$ would modify the kinetic term of Higgs field. Then after canonical normalization, the Higgs mass will also be modified by a factor of $[1+f_{DH}(X_0,\phi_0)]^{-1}$. A simple limit is that the function $f_H(X,\phi)$ which couples to $\mb H^\dag \mb H$ is much smaller than $H^2$, and at the same time, all the rest of $f_\al(X_0,\phi_0)$ in (\ref{InfOSM}) are much smaller than the coefficient of corresponding operator $\O_\al$ in the SM Lagrangian. For instance, if $\O_\al$ is any dimension-4 kinetic term in SM Lagrangian, the previous condition is simply $f_\al(X_0,\phi_0)\ll 1$. In this limit, the masses of SM fields receive negligible amount of correction from inflaton-SM interactions, which presents a universal SM spectrum during inflation.

Finally but not least importantly, we have quantum correction to all ``tree level'' masses considered above. Quantum corrections can be very important when the tree level mass is small. This has been considered in great details in \cite{Chen:2016nrs} and also in the previous section of this paper. We shall include these results in the following when determining the SM spectrum.

With all above points clarified, we can now go to determine the SM spectrum for a general single field slow-roll model assuming that the inflaton is not the SM Higgs field.

\subsection{Higgs Mass}
\label{Sec_HiggsMass}

Firstly we study the Higgs mass $M_{H}^2$. For simplicity we shall call those contributions directly from classical Lagrangian as ``tree mass'', denoted by $M_{H0}^2$, and those from quantum corrections as ``loop mass''. The loop corrected Higgs mass will be denoted by $M_H^2$. We shall firstly consider the various contributions to the tree mass, and then consider the loop corrections.

We are mostly interested in the scenario where the inflation scale is much higher than the electroweak scale so that the tree-level mass term for Higgs field in SM can be well neglected. In this limit, the SM action for Higgs field reduces to (\ref{SHiggs}).
 From now on we parameterize the Higgs doublet in the standard way, $\mb H=\frac{1}{\sqrt 2}(\pi_1+\ii\pi_2, v_h+h+\ii\pi_0)^T$, where $v_h$ is the classical background value of the Higgs field while $\pi_i~(i=0,1,2)$ and $h$ are fluctuation components. As explained above, the expectation value can be classically nonzero in Higgs inflation models which will be studied in Sec.~\ref{Sec_Higgs}, and in the following we will take $v_h=0$.

The tree mass of Higgs field can be nonzero even when the mass term in the classical Lagrangian is set to zero.
This is because, first of all, we can have a dim-4 nonminimal coupling between the Higgs field and the Ricci scalar,
and this introduces a nonzero mass $M_{\xi}^2=12\xi H^2$ in dS background. Besides, we can also consider higher dimensional operators. The most relevant contributions come from the interactions between Higgs field and the inflaton field. Assuming that the shift symmetry of inflaton field in the Lagrangian, we expect that the interactions between Higgs and inflaton have the following form,
\bge
  S\supset-\int\di^4x\sqrt{-g}\Big[f_{H}(X,\phi)\mb{H}^\dag \mb{H}+f_{DH}(X,\phi)|\D_\nu\mb{H}|^2+\cdots\Big],
\ede
in which $f_H(X,\phi)$ and $f_{DH}(X,\phi)$ are aforementioned arbitrary functions of $X= (\pd_\mu\phi)^2$ and $\phi$, and we have neglected operators of higher order in Higgs field $\mb H$ and its derivatives. During inflation the nonzero $\dot\phi_0$ would contribute to the tree mass of Higgs field via the interactions above, and this contribution is given by $\Delta M_{\text{Higgs}}^2=f_{H}(X_0,\phi_0)/[1+f_{DH}(X_0,\phi_0)]$. Note that the factor $f_{DH}(X_0,\phi_0)$ changes the normalization of the Higgs field, thus all tree mass terms are affected by it. In summary, the tree mass of Higgs field during inflation is,
\bge
  M_{H0}^2=\FR{12\xi H^2+f_H(X_0,\phi_0)}{1+f_{DH}(X_0,\phi_0)}.
\ede
The nonminimal coupling $\xi$ is unknown and can be large, and therefore, we should treat the tree mass $M_{H0}^2$ as an input. On the other hand, as discussed above, the most interesting parameter space is the limit $f_{H}(X_0,\phi_0)\ll H^2$ and $f_{DH}(X_0,\phi_0)\ll 1$ so that the inflaton-Higgs coupling would not ``contaminate'' the Higgs mass. We further note that $\xi$, $f_{H}(X_0,\phi_0)$, and $f_{DH}(X_0,\phi_0)$ can be negative in generic EFT. We will mainly consider the parameter region $M_{H0}^2>0$.

Then we need to consider the loop corrections to the Higgs mass. This correction can be very important when $M_{H0}^2\ll H^2$ but is irrelevant in the opposite limit. As explained in \cite{Chen:2016nrs}, this contribution comes from the Higgs loop via $|\mb{H}|^4$  interaction. To quote the result for $O(\mathcal{N})$ theory got in Sec.\;\ref{SecZM}, we can rewrite the SM Higgs doublet as a matrix, $\mb{H}=\frac{1}{\sqrt{2}}h\exp(\ii\pi^i\si^i)$ where $\si^i\;(i=1,2,3)$ are standard Pauli matrices. Then, $\pi^i$ components disappear from the action, and the path integral of $\pi^i$'s zero modes factor out, while the path integral for zero mode of $h$ field has the same form with (\ref{h2vev}) with $\mathcal{N}=4$, as depicted in Sec.\;\ref{SecZM}. Therefore we can evaluate (\ref{h2vev}) at $\mathcal{N}=4$, and use the formula $m_{\text{eff}}^2=\mathcal{N}(V_D\la h^2\ra)^{-1}$ to get,
\bge
\label{HiggsMeff}
M_{H}^2=\sqrt{\FR{\lam}{V_D}}\FR{4\big[1-\sqrt{\pi} ze^{z^2}\text{Erfc}(z)\big]}{-2z+\sqrt{\pi}(1+2z^2)e^{z^2}\text{Erfc}(z)},
\ede
where $\text{Erfc}(z)\equiv 2\pi^{-1/2}\int_z^\infty\di t\,e^{-t^2}$ is the complementary error function, $z= \sqrt{2\pi^2/3\lam}(M_{H0}/H)^2$, and the Higgs self-coupling $\lam$ here is related to its SM value $\lam_{\text{SM}}$ via the following expression due to the presence of inflaton background,
\bge
\lam=\FR{\lam_{\text{SM}}}{[1+f_{DH}(X_0)]^2}.
\ede
The remarkable thing here is that the loop-corrected mass $M_{H}^2$ is nonzero even when the scalar is massless classically ($z=0$) and in this case we have,
\bge
  M_{H}^2=\sqrt{\FR{6\lam}{\pi^3}}H^2.
\ede
This result agrees with \cite{Chen:2016nrs} in qualitative structure $M_{H}^2\sim\sqrt{\lam}H^2$ but differs in the coefficient (with 4 degrees of freedom taken account). The difference can be attributed to the partial resummation of dynamical renormalization group used in \cite{Chen:2016nrs} though we shall not demonstrate this point explicitly in the current work. We plot the quantum corrected mass $M_H$ as a function of tree level mass $M_{H0}$ in Fig.\;\ref{FigScaMass} and it is clear from the plot that the quantum correction dominates when $M_{H0}\ll H$ but becomes negligible quickly when $M_{H0}$ gets larger than $H$.

\begin{figure}[tbph]
\centering
\includegraphics[width=0.5\textwidth]{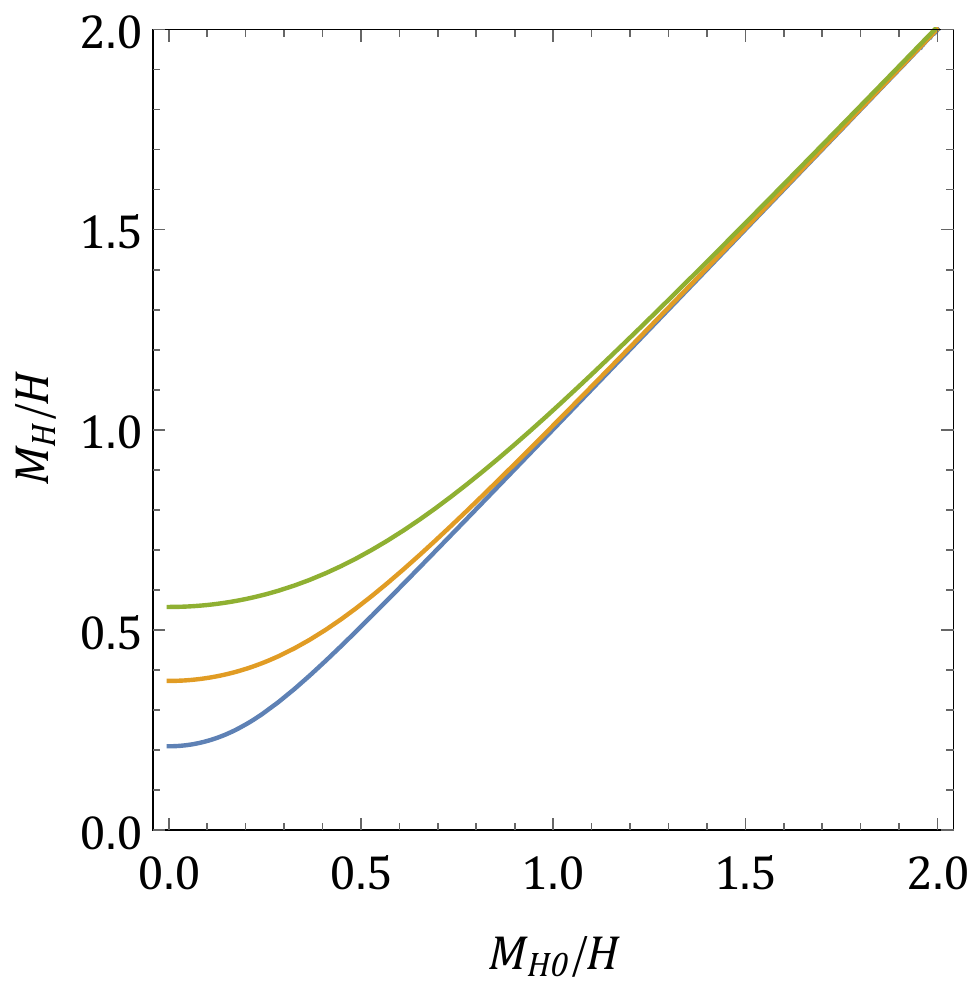}
\caption{The quantum corrected Higgs mass $M_H$ (in unit of Hubble parameter $H$) as functions of tree level mass $M_{H0}$. The three curves from bottom to top correspond to $\lam=(0.01, 0.1, 0.5)$, respectively.}
\label{FigScaMass}
\end{figure}

\subsection{Fermion and Vector Boson Masses}
\label{sec:fvbm}
Gauge bosons and SM fermions do not receive classical mass so long as the gauge symmetry is not broken at tree level, which is the case for our current study since we assume Higgs field does not develop classical VEV during inflation. Therefore, we only need to consider quantum corrections for them. According to our results in the previous section, massless fermions do not receive Dirac mass even at the quantum level, so they remain massless during inflation so long as Higgs does not develop VEV. On the other hand, vector bosons do receive nonzero mass correction from their interactions with scalar fields. The vector bosons in SM include gluon, photon, $W$, and $Z$ bosons. In the unitary gauge, the Higgs field $\mb H$ contains only one real component, the Higgs boson $h$. The gluon and photon do not interact with the Higgs boson $h$ at tree level, so they remain massless during inflation. On the other hand, $W/Z$ bosons can receive nonzero mass due to their interactions with Higgs boson $h$.

Before we can quote the results from previous section, we should also carefully take account of higher dimensional operators involving vector bosons, which could also affect their mass. The most important operators are the following,
\bge
  S\supset-\int\di^4x\sqrt{-g}\Big[f_{DH}(X,\phi)|\D_\mu\mb{H}|^2+\FR{1}{4}f_{W}(X,\phi) W_{\mu\nu}^a W^{\mu\nu a}+\FR{1}{4}f_{B}(X,\phi) B_{\mu\nu}B^{\mu\nu}+\cdots\Big].
\ede
Due to the nonzero background value of inflaton $\phi_0$ and its derivative $\dot\phi_0$, the effective gauge couplings of Higgs field are modified, and are related to their SM values (denoted with subscript ``SM'') in the following way,
\begin{align}
\label{gaugecoup1}
g^2=&~\FR{g_{\text{SM}}^2}{1+f_{W}(X_0,\phi_0)},\\
\label{gaugecoup2}
g'^2=&~\FR{g_{\text{SM}}'^2}{1+f_{B}(X_0,\phi_0)}.
\end{align}

The nonzero quantum corrections to gauge boson mass come again from Higgs loop. We parameterize the Higgs field as a matrix $\mb H=\frac{1}{\sqrt 2}h e^{\ii\pi^i\si^i}$ as was done in Sec.\;\ref{Sec_HiggsMass}, and its covariant derivative is given by $\D_\mu\mb H=\pd_\mu\mb H+\ii g W_\mu^i\si^i\mb H/2-\ii g'B_\mu\mb H\si^3/2$. In the unitary gauge, we set $\pi^i=0$ by gauge rotation, and we can extract interactions between Higgs boson and gauge bosons from the kinetic term of $\mb{H}$ as follows,
\begin{align}
  \FR{1}{2}\tr\big(\D_\mu\mb H^\dag\D^\mu\mb H\big)\supset&~\FR{1}{8}h^2\Big(g^2\sum_{i=1}^3 W_{i \mu}^{} W_i^\mu+g'^2B_\mu B^\mu-2gg'W_{3\mu}B^\mu\Big)\n\\
  =&~\FR{1}{4}g^2h^2\Big(W_\mu^+W^{-\mu}+\FR{1}{2\cos^2\theta_W}Z_\mu Z^\mu\Big).
\end{align}
In the second line above we have switched to the charge and mass eigenbasis, where $W_\mu^\pm=\frac{1}{\sqrt 2}(W_1^\mu\mp\ii W_2^\mu)$, $Z_\mu=W_\mu^3\cos\theta_W-B_\mu\sin\theta_W$, $A_\mu=W_\mu^3\sin\theta_W+B_\mu\cos\theta_W$, and the Weinberg angle $\theta_W$ is given by $\tan\theta_W=g'/g$.
The gauge couplings in the expression above are already modified from their SM values according to (\ref{gaugecoup1}) and (\ref{gaugecoup2}). With these points clarified, we are now ready to write down the quantum corrections to gauge boson masses,
\bge
\label{MW}
\begin{aligned}
  &M_{W}^2=\FR{3g^2H^4}{8\pi^2M_H^2},
  &&M_{Z}^2=\FR{3g^2H^4}{8\pi^2M_H^2\cos^2\theta_W},
\end{aligned}
\ede
and the photon remains massless.

As mentioned in the previous section, so far we have focused on the symmetric phase $M_{H0}^2>0$. Now let us briefly comment the parameter regime $M_{H0}^2<0$, namely $12\xi H^2 + f_H < 0$. In this case, the electroweak symmetry is spontaneously broken and the Higgs field gets a VEV, \footnote{Here we focus on the tree level Higgs potential and neglect the quantum corrections, which should be a good approximation unless $|M_{H0}|\ll H$.}
\begin{align}
  v^2_h = \frac{-4M_{H0}^2}{\lambda}~. 
\end{align}
The tree-level mass of the physical Higgs boson around the potential minima is
\begin{align}
  m_h^2 = \frac{1}{2} \lambda v_h^2 = -2M_{H0}^2~.
\end{align}
This mass should be dominate over the loop-generated mass for small $\lambda$ and $M_{H0}\gtrsim H$. Otherwise the loop (or non-perturbative) corrections should be computed. 

As the case of SM in flat space, the broken electroweak symmetry shifts the mass of $W$ and $Z$ by, \footnote{Strictly speaking, we should recalculate the loop-corrected gauge bosons' mass in the broken phase thus the shift here is just an estimate. But note that the mass shift is dominated by the diagram (\ref{vec2pt}a) instead of diagram (\ref{vec2pt}b). And the interaction in diagram (\ref{vec2pt}a) is not affected by the symmetry breaking. Thus the estimate should be close to the full calculation.}
\begin{align}
  \Delta M_W^2 = \frac{g^2 v_h^2}{4}~,
  \qquad
  \Delta M_Z^2 = \frac{\Delta M_W^2}{\cos^2\theta_W}~.  
\end{align}
The shift can be comparble to the quantum corrections that we have computed above.

More remarkably, all the fermions now get mass from the Higgs mechanism,
\begin{align}
  m_i = \frac{y_i |v_h|}{\sqrt{2}}~, 
\end{align}
where $i$ can be quarks and leptons, namely $t,b,\tau,c,\mu,s,d,u,e$. This does not happen in the case of the symmetric phase $M_{H0}^2>0$. Naturally (considering the $\xi$ parameter), we expect $|v_h|$ is of order $H$, thus the top quark is most likely to be the relevant fermion and leave signatures in cosmology in the case of a broken phase.

\subsection{Renormalization Group Running of SM Coupling Constants}

The SM spectrum in generic non-Higgs inflation studied in this section depends on various SM couplings as well as inflaton-SM couplings. In the case when the contributions from inflaton-SM couplings are negligible, it is possible to make certain predictions to the pattern of SM mass spectrum. For this purpose, we need the Higgs mass, as well as gauge couplings $(g,g')$ associated with $SU(2)_L\otimes U(1)_Y$ as input. Since the inflation scale $H$ can be much higher than the electroweak scale by several orders of magnitude, the renormalization group (RG) running of gauge couplings should be taken into account. Furthermore, in the case when fermions do acquire mass during inflation, the RG running of Yukawa coupling is important, too.

More broadly speaking, the Higgs self-coupling $\lam$ is perhaps the most crucial part of RG running of SM couplings in the context of inflation. Indeed, current measurements of Higgs mass and top mass suggest that the $\be$ function for $\lam$ is negative, and that $\lam$ would probably turn negative at some high energy scale $\mu_0$. Using SM 2-loop RG running together with the current central values of Higgs and top mass, one can find $\mu_0\sim 10^{11}$GeV, which is indeed possible to be lower than the inflation scale (measured by Hubble parameter). Meanwhile, there are large uncertainties in this calculation, chiefly because the turning scale $\mu_0$ is exponentially sensitive to the input of Higgs mass and top mass at electroweak scale. Consequently, the uncertainties associated with the Higgs and top mass measurement would greatly affect the prediction of $\mu_0$. In addition to that, the calculation of effective potential usually suffers from the problem of gauge dependence. Different treatment of this issue would also affect the result \cite{Andreassen:2014eha,Andreassen:2014gha}.

Further complication appears when one considers all these problems during inflation, where a lot more factors need to be considered, e.g. the nonminimal coupling between Higgs field and Ricci scalar, the stochastic quantum fluctuation of Higgs field. One should also be concerned with the Higgs instability during reheating epoch even this instability is not that harmful during inflation \cite{Hook:2014uia,Herranen:2014cua,Kearney:2015vba,Espinosa:2015qea}.

We shall not consider these problems further in current work, but only take the attitude that there \emph{is} a successful inflation scenario which correctly generates the density perturbation as we see today, and is free from the problem of Higgs instability. Perhaps the simplest way to achieve this scenario is to assume some new physics beyond SM to stabilize the Higgs sector. Without further digression, now we show the inflation scale dependence of SM spectrum by applying 2-loop RG running. The two-loop $\beta$ functions for gauge sector, Higgs self-coupling, and top-Yukawa coupling, with non-minimal coupling $\xi$ included, can be found in \cite{Allison:2013uaa}, the two-loop $\beta$ functions for all Yukawa couplings can be found in \cite{Bednyakov:2014pia}, and the 1-loop matching conditions can be found in \cite{Degrassi:2012ry}. We identify the running scale to be the Hubble scale $H$, and then plot the gauge boson masses (\ref{MW}), and Yukawa couplings $y_i$, as functions of $H$, in Fig.\;\ref{FigRG}. In the most optimistic scenario where we can measure the gauge boson mass splitting and the corresponding amplitudes accurately, it is possible to obtain the scale of inflation $H$ from such a measurement, which is rather difficult because it requires us to firstly identify the signals of gauge bosons. One possible way to achieve this goal is to try to identify the consistency relation for gauge boson signals (\ref{eq:cons-rel}) as will discussed in the following. Given all practical difficulties for such observations, this is however an interesting point to make because it means that we may be lucky to learn the scale of inflation solely from scalar mode of primordial perturbation due to the renormalization group running of the particle spectrum, without any reference to the tensor mode.

\begin{figure}[tbph]
\centering
\includegraphics[height=50mm]{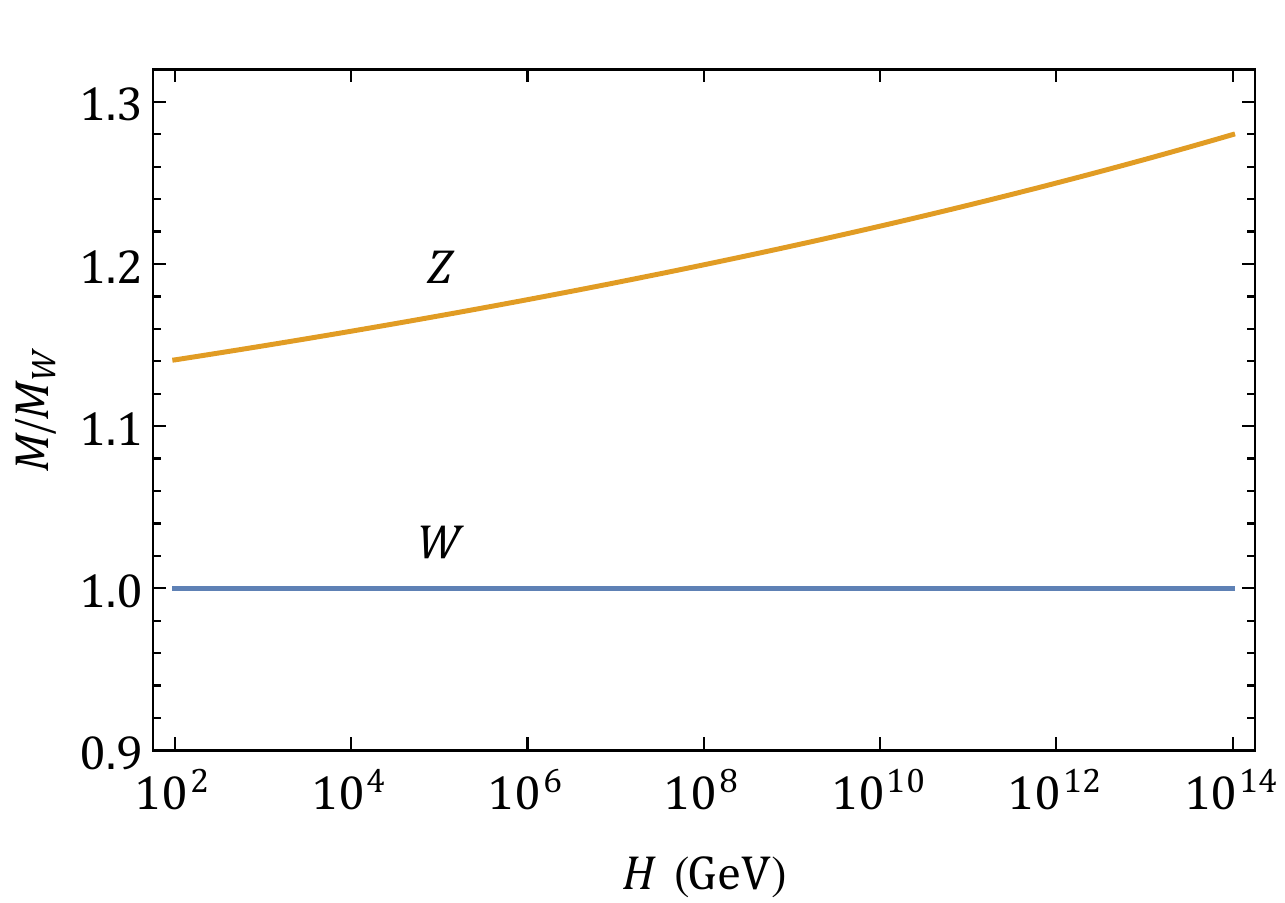}
\includegraphics[height=50mm]{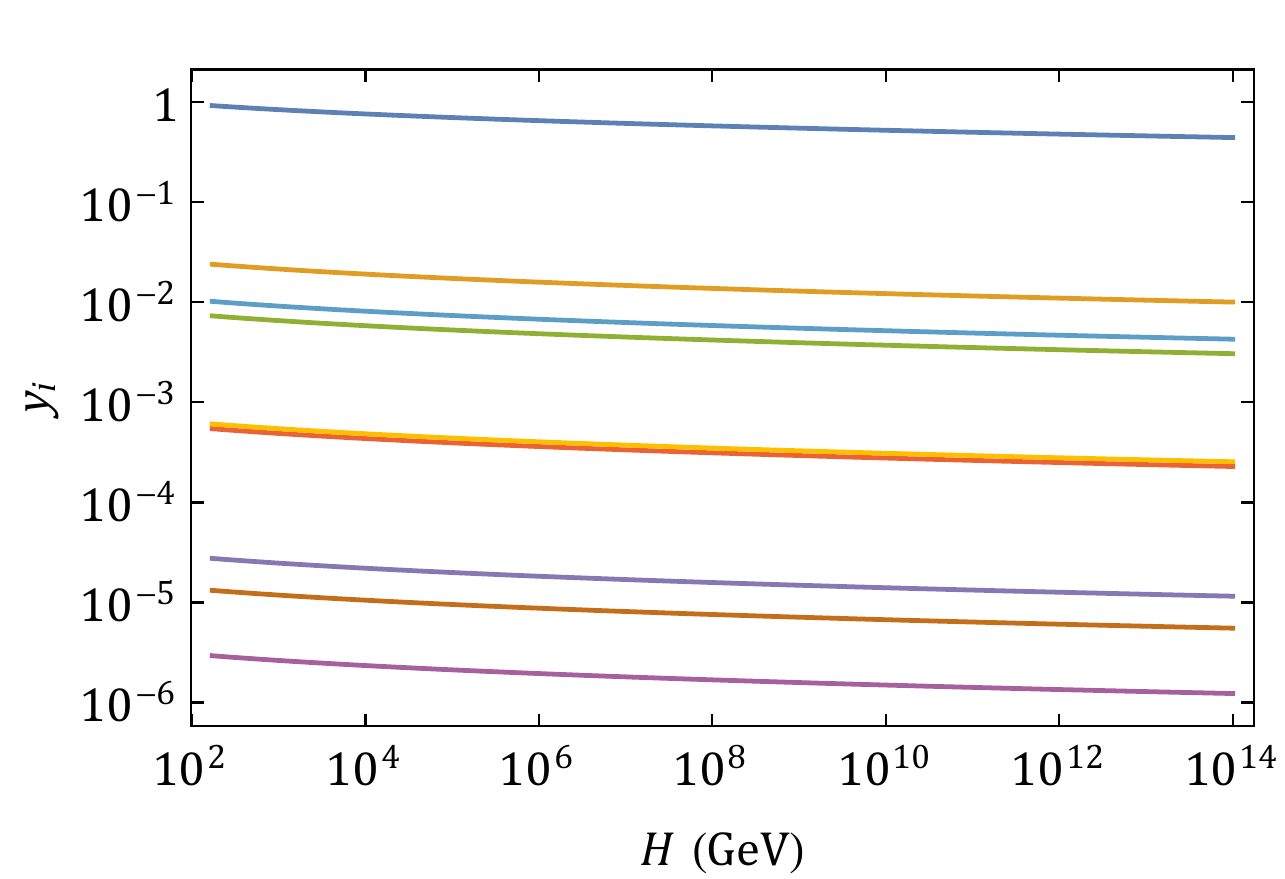}
\caption{ The gauge boson masses (left panel) and Yukawa couplings (right panel) as functions of Hubble scale $H$, using 2-loop SM renormalization group running. On left panel, the masses of $Z$ and $W$ (from top to bottom) are normalized by the mass of $W$; on right panel, various curves from top to bottom correspond to Yukawa couplings of $t,b,\tau,c,\mu,s,d,u,e$, respectively.}
\label{FigRG}
\end{figure}

\section{Three-Point Function in Non-Higgs Inflation}
\label{Sec_3pt}

In previous sections we have worked out the mass spectrum of SM particles during inflation. It remains to answer how the SM spectrum can be probed observationally. It has been known that a massive field can reveal itself during inflation through characteristic scaling behaviors in the squeezed limit of bispectrum of the primordial curvature perturbation \cite{Chen:2009we,Chen:2009zp,Baumann:2011nk,Assassi:2012zq, Noumi:2012vr, Arkani-Hamed:2015bza}, due to its interaction with the inflaton field. In the case of SM, if we assume that the inflaton is SM singlet, then the SM fields can couple to the inflaton field only through singlet operators. This implies in particular that SM fields appear in its inflaton coupling at least pairwise, and that SM fields can contribute to inflaton three-point function starting at 1-loop level. So it is again important to study SM loops during inflation.

In this section we are going to evaluate the squeezed limit of bispectrum with intermediate loops of SM fields. We shall assume that the inflaton field $\phi$ interacts with SM fields through the action (\ref{InfOSM}). Expanding the inflaton field $\phi=\phi_0+\de\phi$ around its background value $\phi_0$, we get the leading couplings between inflaton fluctuation $\de\phi$ and SM fields to be,
\begin{align}
\label{fOexpand}
  -\int\di^4x\sqrt{-g} \sum_\al f_\al(X,\phi)&~\O_\al\n\\
  =-\int\di^4x\sqrt{-g} \sum_\al \Big[&f_\al(X_0,\phi_0)+f_{\al,\phi}(X_0,\phi_0)\de\phi-2f_{\al,X}(X_0,\phi_0)\dot\phi_0\dot{\de\phi}\n\\
  &+\FR{1}{2}f_{\al,\phi\phi}(X_0,\phi_0)(\de\phi)^2-2f_{\al,X\phi}(X_0,\phi_0)\dot\phi_0\dot{\de\phi}\de\phi\n\\
  &+f_{\al,X}(X_0,\phi_0)(\pd_\mu\de\phi)^2+2f_{\al,XX}(X_0,\phi_0)\dot\phi_0^2(\dot{\de\phi})^2\Big]\O_\al,
\end{align}
in which $X$ and $\phi$ in subscripts of $f_\al$ functions after a comma denote corresponding derivatives of $f_\al$. We have kept all terms up to quadratic order in fluctuation field $\de\phi$. Some further simplification can be made for our current calculation. Firstly, we shall postpone the study of direct coupling to Sec.\,\ref{Sec_Higgs}, and here we shall assume that all $f_\al$ functions depend only on $X$ but not directly on $\phi$. In other words, we shall consider the leading terms in the approximate shift symmetry. Secondly, we shall keep only the leading terms in the slow-roll approximation, so that terms suppressed by more powers of $\dot\phi_0$ can be dropped off. Under these two assumptions\footnote{The latter assumption is not quite robust. For instance, in the case $f_\al(X)\sim X^2$, we see that the ``$\dot\phi_0^2$-suppressed'' term $f_{\al,XX}(X_0)\dot\phi_0^2(\dot{\de\phi})^2$ is actually of the same order as $f_{\al,X}(X_0)(\pd_\mu\de\phi)^2$. We can nevertheless keep all such terms and the calculation in the following would be almost the same, although the expression would be a little more complicated.}, there are only 3 terms left in (\ref{fOexpand}),
\begin{align}
\label{fOExpLead}
  &-\int\di^4x\sqrt{-g} \sum_\al f_\al(X,\phi)\O_\al\n\\
  \supset&-\int\di^4x\sqrt{-g} \sum_\al \Big[f_\al(X_0)-2f_{\al}'(X_0)\dot\phi_0\dot{\de\phi}+f_{\al}'(X_0)(\pd_\mu\de\phi)^2\Big]\O_\al,
\end{align}
where we have dropped off the explicit $\phi$ dependence of $f_\al$, and used a prime to denote the derivative of $f_\al$ with respect to $X$. In (\ref{fOExpLead}), the first term proportional to $f_\al(X_0)$ is a rescaling of the corresponding operator $\O_\al$. As elaborated in last section, this term shall modify the SM spectrum significantly unless it is sufficiently small.

We are mostly interested in the case that $\mathcal{O}_{\al}$ is quadratic in SM fields because only such operators can contribute 1-loop diagrams. We shall consider operators with mass dimension up to 4, then the only choices for $\O_\al$ are the following,
\begin{align}
\label{OSM}
S\supset -\int\di^4x\sqrt{-g}\Big[&\,f_H(X)\mb H^\dag\mb H+f_{DH}(X)|\D_\mu \mb H|^2-f_{\Psi_i}(X)\ii\ob{\Psi}_i\sla{\mathscr{D}}\Psi_i\n\\
&+\FR{1}{4}f_{W}(X)W^{}_{a\mu\nu}W_a^{\mu\nu}+\FR{1}{4}f_{B}(X)B^{}_{\mu\nu}B^{\mu\nu}\Big].
\end{align}
Most of them have appeared in the previous section, except the one for fermions. Though fermions remain massless in non-Higgs inflation scenarios, we still consider massive fermions in this section for completeness, and the result will also be useful in Sec.\,\ref{Sec_Higgs}.

In this paper we only consider $\O_\al$ of spin-0, i.e., scalar operators. It is well possible that the SM fields couple to inflaton via higher spin interactions, such as $(\pd_{\mu}\pd_\nu\phi)(\pd^\mu\phi)\ob{\Psi}_i\ga^\nu\Psi_i$. Such higher spin operator will leave characteristic angular dependence on the bispectrum \cite{Arkani-Hamed:2015bza}. The important point here is that a detection of such higher-spin behavior does not necessarily imply the discovery of a corresponding higher spin particle, because it is possible that it is only a higher spin superposition of some lower spin particles running in the loop. For example, the operator $(\pd^\mu\pd^\rh\phi)(\pd^\nu\pd^\si\phi)F_{\mu\nu}F_{\rh\si}$ can generate a spin-2 angular dependence in the bispectrum of $\phi$, although this is from the superposition of a pair of spin-1 boson, not from a graviton.

Using the Schwinger-Keldysh formalism, the 3-point correlator of inflaton fluctuation $\de\phi$ with 1-loop contribution from operator $\O_\al$ can be written as,
\begin{align}
\label{3ptInt}
&~\big\la\de\phi(\mb k_1)\de\phi(\mb k_2)\de\phi(\mb k_3)\big\ra_{\al}'\n\\
=&~4f_\al'^{\,2}(X_0) \sum_{a,b=\pm}ab\int_{-\infty}^0\FR{\di\tau'}{(H\tau')^2}\int_{-\infty}^0\FR{\di\tau''}{(H\tau'')^2}\Big[-\pd_{\tau''}G_{+b}(\mb k_3,\tau,\tau'')\pd_{\tau''}\phi_0\Big]
\n\\
&~\times\Big[-\pd_{\tau'}G_{+a}(\mb k_1,\tau,\tau')\pd_{\tau'}G_{+a}(\mb k_2,\tau,\tau')-\mb k_1\cdot\mb k_2 G_{+a}(\mb k_1,\tau,\tau')G_{+a}(\mb k_2,\tau,\tau')\Big]\n\\
&~\times \int\di^3X\,e^{-\ii\mb k_I\cdot \mb X}\Big\la \mathcal{O}_{\al}(\tau',\mb x')\mathcal{O}_{\al}(\tau'',\mb x'')\Big\ra_{ab}\n\\
=&~\FR{f_\al'^{\,2}(X_0)H\dot\phi_0}{2k_1^3k_2^3k_3^3}\n\\
 &~\times\sum_{a,b=\pm}ab\int_{-\infty}^0\FR{\di\tau'}{\tau'^2}\int_{-\infty}^0\FR{\di\tau''}{\tau''^2}k_3^2\big[-k_1^2k_2^2\tau'^2-\mb k_1\cdot\mb k_2(1-\ii a k_1\tau')(1-\ii a k_2\tau')\big]\n\\
&~\times e^{\ii a(k_1+k_2)\tau'+\ii b k_3\tau''}\int\di^3X\,e^{-\ii\mb k_I\cdot \mb X}\Big\la \mathcal{O}_{\al}(\tau',\mb x')\mathcal{O}_{\al}(\tau'',\mb x'')\Big\ra_{ab}\n\\
\equiv&~\FR{f_\al'^{\,2}(X_0)H\dot\phi_0}{2k_1^3k_2^3k_3^3}\mathcal{I}(k_1,k_2,k_3),
\end{align}
where $\mb k_I=\mb k_1+\mb k_2$, $\mb X\equiv \mb x'-\mb x''$, and the indices $(a,b)$ take either plus or minus sign, which correspond to the $+$ and $-$ contour in Schwinger-Keldysh formalism. More details about Schwinger-Keldysh formalism and Feynman rules during inflation used in this paper can be found in \cite{Chen:2016nrs}. The expectation value $\la\O_\al^2\ra$ is in general rather difficult to work out, and it is even more difficult to carry out the whole integral.

Fortunately, under certain approximations, it is possible to get analytic expressions for the part of amplitude which we are mostly interested in. The first approximation we shall take, is to expand the 2-point correlator $\la\O_\al(\tau',\mb x')\O_\al(\tau'',\mb x'')\ra$ in the late time (IR) limit $\tau',\tau''\to 0$. Note that, while this is a good approximation for the massive fields at the vertex point $\tau'$ (at which $k_{1,2}\simeq k_{\rm {massless}} \gg k_{\rm {massive}}=k_3$), it is not always a good approximation for the massive fields at the vertex point $\tau''$ (at which $k_{\rm {massless}}=k_{\rm {massive}}=k_3$). In the latter case, it is a reasonable approximation for $\mu\sim 1$ but not for $\mu\gg 1$. This is because the IR approximation of the massive field mode function is good for $0<k_{\rm {massive}}|\tau|<\sqrt{\mu}$.  The interaction between the massive and massless mode (such as the resonance) takes effect around $k_{\rm {massless}}|\tau|=\mu$. If $k_{\rm {massless}} \sim k_{\rm {massive}}$, the interaction point is close to the validity region of the IR approximation $0<k_{\rm {massive}}|\tau|<\sqrt{\mu}$ only if $\sqrt{\mu}\sim 1$ but getting worse for $\mu \gg 1$. On the other hand, if $k_{\rm {massless}}/k_{\rm {massive}} > \sqrt{\mu}$, the interaction point lies in the validity region of the IR approximation. In any case, the $\mu\sim 1$ case is sufficient for our purpose because more massive fields contribute Boltzmann factors and become less interesting phenomenologically.


The second simplifying assumption we shall make, is that we are only concerned with the so-called nonlocal part of the amplitude, i.e., the part of the momentum-dependence proportional to $(k_3/k_{1})^\ga$ with $\ga$ being some real (generically non-integer) or complex number. It is this part of the power-law or oscillatory behavior in momentum ratio that encodes the mass spectrum of particle states. Keeping nonlocal part only is also a desirable simplification because, as we shall see below, for generic values of mass, the nonlocal part of the expectation value is disentangled with the UV divergence of the loop integral, and therefore, we do not have to run into the problem of regularization and renormalization, which is a notoriously difficult task in dS. Moreover, the nonlocal part of the late time expansion of $\la\O_\al^2\ra$ is actually independent of $\pm$ contour of in-in integral, and for this reason we can freely drop the $ab$ indices of the 2-point correlator $\la\O_\al^2\ra$.

We have one more computational simplification as pointed out in \cite{Arkani-Hamed:2015bza}. That is, the amplitude (\ref{3ptInt}) can be easily got by firstly computing a much simpler 4-point amplitude of conformal scalars $\phi_c$, and then applying a differential operator on the amplitude of conformal scalars. Assuming that the conformal scalar $\phi_c$ couples to $\O_\al$ through direct coupling $\int\di^4x\sqrt{-g}\phi_c^2\O_\al$, and making use of the following mode function of a conformal scalar $\phi_c$ in dS,
\bge
  \phi_c(\tau,\mb k)=\FR{H\tau e^{-\ii k\tau}}{\sqrt{2 k}},
\ede
we can write down a 4-point amplitude of the conformal scalar $\phi_c$ contributed by the $\la\O_\al^2\ra$ as,
\begin{align}
\label{4ptInt}
&~\sum_{a,b=\pm}ab\int_{-\infty}^\tau\FR{\di\tau'}{(H\tau')^4}\int_{-\infty}^\tau\FR{\di\tau''}{(H\tau'')^4}G^{(c)}_{+a}(\mb k_1;\tau,\tau')G^{(c)}_{+a}(\mb k_2;\tau,\tau')G^{(c)}_{+b}(\mb k_3;\tau,\tau'')G^{(c)}_{+b}(\mb k_4;\tau,\tau'')\n\\
  &\times\int\di^3X\,e^{-\ii\mb k_I\cdot \mb X}\Big\la \mathcal{O}_{\al}(\tau',\mb x')\mathcal{O}_{\al}(\tau'',\mb x'')\Big\ra_{ab}\n\\
=&~\FR{4\tau^4}{16k_1k_2k_3k_4}\sum_{a,b=\pm}ab\int_{-\infty}^\tau\FR{\di\tau'}{\tau'^2}\int_{-\infty}^\tau\FR{\di\tau''}{\tau''^2}e^{\ii a(k_1+k_2)\tau'+\ii b(k_3+k_4)\tau''}\n\\
&~\times\int\di^3X\,e^{-\ii\mb k_I\cdot \mb X}\Big\la \mathcal{O}_{\al}(\tau',\mb x')\mathcal{O}_{\al}(\tau'',\mb x'')\Big\ra\n\\
\equiv&~\FR{4\tau^4}{16k_1k_2k_3k_4}\mathcal{I}_c(k_{12},k_{34},k_I),
\end{align}
where $G^{(c)}_{ab}$ is the propagator of conformal scalar; $k_{12}=k_1+k_2$ and $k_{34}=k_3+k_4$, It is clear that the integrals $\mathcal{I}(k_1,k_2,k_3)$ in (\ref{3ptInt}) and $\mathcal{I}_c(k_{12},k_{34},k_I)$ in (\ref{4ptInt}) have similar form when $\tau\to 0$, apart from the complicated factor involving $\mb k_1$ and $\mb k_2$ in $\mathcal{I}(k_1,k_2,k_3)$. This difference is inessential as was pointed out in \cite{Arkani-Hamed:2015bza}, and can be removed by acting an differential operator on $\mathcal{I}_c(k_{12},k_{34},k_I)$ in the following way,
\begin{align}
\label{O12}
\mathcal{I}(k_1,k_2,k_3)
=k_3^2\Big[k_1^2k_2^2\pd_{k_{12}}^2-\mb k_1\cdot\mb k_2(1-k_{12}\pd_{k_{12}}+k_1k_2\pd_{k_{12}}^2)\Big]\mathcal{I}_c(k_{12},k_3,k_3).
\end{align}

Finally, we comment on the evaluation of $\la\O_\al^2\ra$. Since we are only concerned with 1-loop correction, the only relevant $\O_\al$ consists of operators quadratic in SM fields. Therefore, the expectation value $\la\O_\al(x)\O_\al(x')\ra$ is essentially the same with a four-point correlation of SM fields $\Phi$, i.e. $\la\Phi(x_1)\Phi(x_2)\Phi(x_3)\Phi(x_4)\ra$. At leading order, we have three diagrams contributing to this correlation, namely the usual $s$, $t$, $u$ channels. However, as the four points $x_i$ are identified pairwise, $x_1=x_2=x$ and $x_3=x_4=x'$, the $t$-channel contribution will be divergent, and it is actually a part of definition of $\O_\al$ that this divergence should be subtracted. As a result, only $s$ channel and $u$ channel diagrams are left, as shown diagrammatically below.
\bge
\parbox{0.75\textwidth}{\vspace{0mm}\includegraphics[width=0.75\textwidth]{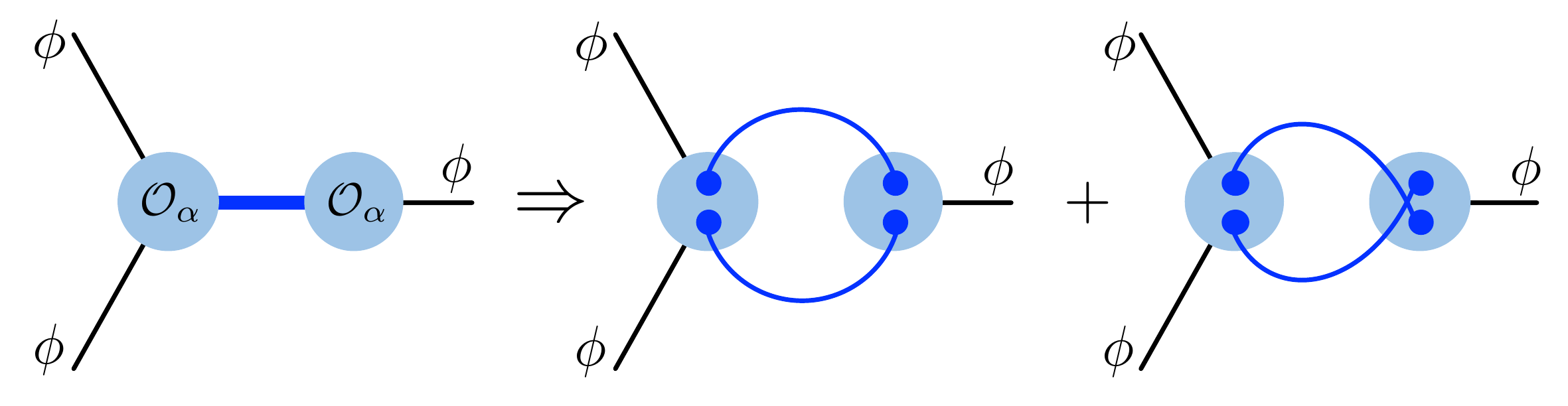}}
\ede
In the case of fermion where the two fields in $\O_\al$ are not identical ($\psi$ and $\ob\psi$), the $u$ channel is absent too, and in that case only one fermion loop contributes.

With above technical fine points clarified, we are now ready to evaluate the expectation value $\la\O_\al^2\ra$ for all operators in (\ref{OSM}).

\subsection{$\mb H^\dag\mb H$}

The process of calculating three-point function as done above applies to each of the operators in (\ref{OSM}) given the late-time expanded form of $\la\O_\al^2\ra$. Therefore, we shall not present all the intermediate steps of calculation for these operators. Instead, we shall explain how the late-time expansion of $\la\O_\al^2\ra$ is calculated in each case.

We first consider the unique dimension-2 gauge singlet operator,
\begin{align}
  S\supset &-\FR{1}{1+f_{DH}(X_0)}\int\di^4x\sqrt{-g}f_{H} (X)\mb{H}^\dag\mb{H}\n\\
  =&-\FR{1}{2\big(1+f_{DH}(X_0)\big)}\int\di^4x\sqrt{-g}f_{H} (X)h^2,
\end{align}
where we have kept the only physical degree of freedom $h$ in the Higgs doublet $\mb H$, and we have included a factor of $[1+f_{DH}(X_0)]^{-1}$ so that the Higgs field is canonically normalized in the presence of the inflaton background.
Then according to the aforementioned strategy, we need to work out the late time expansion of the following expectation value,
\begin{align}
&\Big\la\big[\mb H^\dag\mb H(x)\big]\big[\mb H^\dag\mb H(x')\big]\Big\ra=\FR{1}{4}\big\la h^2(x)h^2(x')\big\ra=\FR{1}{2}\big\la h(x)h(x')\big\ra^2=\FR{1}{2}G_h^2(x,x').
\end{align}
The propagator $G_h(x,x')$ of $h$ field can be expanded at late time limit, using (\ref{Zinf}), as,
\begin{align}
  G_h(x,x')
  =&~\FR{H^2}{16\pi^2}\Gamma(\FR{3}{2}-\mu_h)\Gamma(\FR{3}{2}+\mu_h)\,{}_2F_1\Big(\FR{3}{2}-\mu_h,\FR{3}{2}+\mu_h;2;\FR{1+Z_{xx'}}{2}\Big)\n\\
  \To&~\FR{H^2}{4\pi^{5/2}}\bigg[\Gamma(\mu_h)\Gamma(\FR{3}{2}-\mu_h)\Big(\FR{\tau\tau'}{X^2}\Big)^{3/2-\mu_h}+(\mu_h\to-\mu_h)\bigg],
\end{align}
where $\mu_h=\sqrt{9/4-(M_H/H)^2}$, and $Z_{xx'}$ is the embedding distance between $x$ and $x'$, defined in Appendix \ref{AppEdS1}. Therefore, we get
\begin{align}
 &G_h^2(x,x')
=\FR{H^4}{16\pi^5}\bigg[\Gamma(\mu_h)^2\Gamma(\FR{3}{2}-\mu_h)^2\Big(\FR{\tau\tau'}{X^2}\Big)^{3-2\mu_h}+(\mu_h\to-\mu_h)\bigg],
\end{align}
and its Fourier transformation,
\begin{align}
  &\int\di^3X\, e^{-\ii\mb k_I\cdot \mb X}G_h^2(x,x')\n\\
  =&~\FR{H^4(\tau\tau')^{3/2}}{4\pi^4}\bigg[\Gamma(\mu_h)^2\Gamma(\FR{3}{2}-\mu_h)^2\Gamma(-4+4\mu_h)\sin(2\pi\mu_h)(k_I^2\tau\tau')^{3/2-2\mu_h} + (\mu_h\to-\mu_h)\bigg].
  \label{Fourier_trans}
\end{align}
In these expression we have kept the nonlocal terms only, that is, terms that are not the polynomials of $k_I^2$. According to previous discussion, only such terms can contribute to characteristic power-law/oscillatory behavior. Then we can use (\ref{3ptInt}) to evaluate the three-point function of the inflaton perturbation. Remarkably, each of four in-in integral in (\ref{3ptInt}) is divergent in the late time region, but the divergences cancel out among the four, as they should. The finite result in the squeezed limit $k_{1,2}\gg k_3$ can then be written as follows,
\begin{align}
\label{3ptH2}
  &\big\la\de\phi(\mb k_1)\de\phi(\mb k_2)\de\phi(\mb k_3)\big\ra_{H}\n\\
  &=-\bigg[\FR{f_{H}' (X_0)}{1+f_{DH}(X_0)}\bigg]^2\FR{H^5\dot\phi_0}{8\pi^4k_S^6}\bigg[C_{H}(\mu_h)\Big(\FR{k_L}{2k_S}\Big)^{-2\mu_h}+(\mu_h\to-\mu_h)\bigg],
\end{align}
where we have made the approximation $k_1\simeq k_2\equiv k_S$ (and therefore, $k_{12}\simeq 2k_S$) as well as $k_3\equiv k_L$, and the $\mu$-dependent coefficient $C_{H}(\mu)$ is given by,
\bge
C_{H}(\mu)\equiv(2-\mu)(3-2\mu)\cos(\pi\mu)\sin^3(\pi\mu)\Gamma(-4+4\mu)\Gamma^2(\mu)\Gamma^2(\FR{3}{2}-\mu)\Gamma^2(2-2\mu).
\ede

When $M_H>\frac{3}{2}H$, $\mu_h$ is pure imaginary, and the two terms in (\ref{3ptH2}) are complex conjugate of each other. One can readily see the Boltzmann suppression factor $e^{-2\pi M_H/H}$ from $C_H(\mu)$ when $M_H/H\gg 1$, by using the Stirling expansion $\Gamma(z)\sim
\sqrt{2\pi}e^{-z}z^{z-1/2}$ when $z\to \infty$. In this case we have $\mu\sim \ii M_H/H$, and for finite real $a$ and $b$, we have $\Gamma(a+b \mu)\sim \Gamma(\ii b M_H/H)\sim e^{-\pi |b| M_H/2H}/\sqrt{|b|M_H/H}$. So all $\Gamma$ functions in $C_H(\mu)$ contribute $e^{-6\pi M_H/H}$ in total, together with another factor of $e^{+4\pi M_H/H}$ coming from trigonometric functions, we see that the Boltzmann suppression factor $e^{-2\pi M_H/H}$ is recovered in the large $M_H/H$ limit. As expected, this suppression is the square of the tree-level case because two massive fields are excited quantum-mechanically in the loop diagram.

On the other hand, when Higgs mass $M_H< \frac{3}{2}H$, we have $\mu_h>0$, and the term explicitly shown in (\ref{3ptH2}) dominates the squeezed limit. We note that the amplitude \eqref{3ptH2} has poles when $\mu_h=1/4, 3/4, 3/2$. The poles at $\mu_h=1/4,3/4$ are unphysical and are due to the UV divergence arising from \eqref{Fourier_trans}. Although, as mentioned, for generic values of $\mu_h$ the UV divergence is absent after several approximation methods, for some special values of $\mu_h$ it is still present. We leave the more proper treatment for future investigation. The presence of the pole at $\mu_h=3/2$ is however expected, because in this case the Higgs becomes massless and contribute to the correlation indefinitely at super-horizon scales. An infrared cutoff, such as the end of inflation, or a dynamically built mass, should be present to regulate the pole \cite{Chen:2009zp}.


The above discussions of imaginary $\mu_h$ and $\mu_h>0$ regimes also apply for the subsections that follow, where similar coefficients arise for other types of operators.

\subsection{$|\D_\mu\mb H|^2$}

Now we consider the other operator $\O_\al=|\D_\mu \mb H|^2$ involving Higgs field in (\ref{OSM}),
\bge
  S\supset-\FR{1}{1+f_{DH}(X_0)}\int\di^4x\sqrt{-g}f_{DH} (X)|\D_\mu \mb{H}|^2.
\ede
Here again we include the normalization factor $[1+f_{DH}(X_0)]^{-1}$ so that $\mb H$ is properly normalized. We again extract the only physical component $h$ from $\mb H$. Then we need to evaluate the following 2-point correlation function,
\begin{align}
\Big\la\big[(\pd_\mu h)^2(x)\big]\big[(\pd_\nu h)^2(x')\big]\Big\ra
=2\big[\nabla_\mu\nabla_{\nu'} G_h(x,x')\big]\big[\nabla^\mu\nabla^{\nu'}G_h(x,x')\Big].
\end{align}
We have rewrite partial derivates as covariant derivatives since they are equivalent for scalar functions. Then the covariant derivatives of the propagator $G_h(Z)$ as a function of imbedding distance $Z=Z(x,x')$ can be worked out using the formulae in Appendix \ref{AppEdS}, as follows,
\begin{align}
&2\big[\nabla_\mu\nabla_{\nu'} G_h(Z)\big]\big[\nabla^\mu\nabla^{\nu'}G_h(Z)\Big]\n\\
=&~6H^4 G_h'^2(Z)+2H^4\big[(1-Z^2)G_h''(Z)-Z G_h'(Z)\big]^2
\end{align}
Then we are ready to expand the propagator and its derivatives at late time limit $\tau,\tau'\to 0$, using (\ref{Zinf}), to get a pair of nonlocal terms at leading order,
\begin{align}
&\Big\la\big[(\pd_\mu h)^2(x)\big]\big[(\pd_\nu h)^2(x')\big]\Big\ra\n\\
=&~\FR{H^8}{32\pi^5}\bigg[\Gamma^2(\mu_h)\Gamma^2(\FR{5}{2}-\mu_h)(3-2\mu_h)^2\Big(\FR{\tau\tau'}{X^2}\Big)^{3-2\mu_h}+(\mu_h\to-\mu_h)\bigg].
\end{align}
Once we have this expanded form of 2-point correlation, we can proceed directly to calculate the following three-point function,
\begin{align}
  &\big\la\de\phi(\mb k_1)\de\phi(\mb k_2)\de\phi(\mb k_3)\big\ra_{DH}\n\\
  &=-\bigg[\FR{f_{DH}' (X_0)}{1+f_{DH}(X_0)}\bigg]^2\FR{H^9\dot\phi_0}{32\pi^4k_S^6}\bigg[C_{DH}(\mu_h)\Big(\FR{k_L}{2k_S}\Big)^{-2\mu_h} +(\mu_h\to-\mu_h) \bigg],
\end{align}
where
\begin{align}
  C_{DH}=&~(2-\mu_h)(3-2\mu_h)^3\cos(\pi\mu_h)\sin^3(\pi\mu_h)\n\\
  &~\times\Gamma(-4+4\mu_h)\Gamma^2(\mu_h)\Gamma^2(\FR{5}{2}-\mu_h)\Gamma^2(2-2\mu_h),
\end{align}
and $k_L,k_S$ are defined below (\ref{3ptH2}).
This amplitude has unphysical poles at $\mu_h=1/4,3/4$.

\subsection{$\ob{\Psi}\ii\sla{\mathscr{D}}\Psi$}
For the Dirac spinor of mass $M_F$ we consider the following interaction term with the spinor field $\Psi_i$ properly normalized,
\bge
  S\supset \FR{1}{1+f_{\Psi }(X_0)}\int\di^4x\sqrt{-g}f_{\Psi }(X)\ii\ob{\Psi} \ga^\mu\mathscr{D}_\mu\Psi .
\ede
The corresponding 2-point correlation we need to calculate is the following,
\begin{align}
\label{spinorloop2pt}
&\Big\la\big[\ob{\Psi}\sla{\nabla}\Psi(x)\big]\big[\ob{\Psi}\sla{\nabla}\Psi(x')\big]\Big\ra\n\\
=&~M_F^2\Big\la\big[\ob{\Psi}\Psi(x)\big]\big[\ob{\Psi}\Psi(x')\big]\Big\ra\n\\
=&-M_F^2\tr\Big[G_F(x,x')G_F(x',x)\Big],
\end{align}
in which we have used the Dirac equation in the first equality. Since we are now interested in the late time behavior of this quantity, it is more convenient to work in real time dS rather than doing Wick rotation. The propagator of a massive Dirac fermion in $dS_D$ is well-known and given by \cite{Cande1975,Allen:1986qj,Miao2006,Koksm2009,Miao2012},
\bge
G_F(x,x')=H^{D-2} a(x)(\ii\sla\nabla+M_F)\Big[S_+(x,x')\FR{1+\ga^0}{2}+S_-(x,x')\FR{1-\ga^0}{2}\Big],
\ede
where
\begin{align}
  S_\pm(x,x')=&~\FR{1}{(4\pi)^{D/2}\sqrt{a(x)a(x')}}\FR{\Gamma(\frac{D}{2}-1\mp\ii {\mu_{1/2}})\Gamma(\frac{D}{2}\pm\ii {\mu_{1/2}})}{\Gamma(\frac{D}{2})}\n\\
  &~\times{\,}_{2}F_1\Big(\FR{D}{2}-1\mp\ii {\mu_{1/2}},\FR{D}{2}\pm\ii {\mu_{1/2}};\FR{D}{2};\FR{1+Z}{2}\Big),
\end{align}
in which ${\mu_{1/2}}\equiv M_F/H$, and $Z=Z(x,x')$ is the imbedding distance between $x$ and $x'$. From now on we shall set $D=4$ since no regularization is needed for nonlocal part of late time expansion of (\ref{spinorloop2pt}). For notational simplicity we define $A_\pm(x,x') =\frac{1}{2}[S_+(x,x') \pm S_-(x,x')]$. Note that $A_\pm(x,x')$ is symmetric with respect to its two arguments, so we shall not write the arguments explicitly. Then (\ref{spinorloop2pt}) can be evaluated as follows,
\begin{align}
  &-M_F^2\tr\Big[G_F(x,x')G_F(x',x)\Big]\n\\
 =&-H^4M_F^2a(x)a(x')\tr\Big\{(\ii\sla\nabla_\mu+M_F)(A_++A_-\ga^0)(\ii\sla\nabla_{\mu'}+M_F)(A_+ +A_- \ga^0)\Big\}\n\\
 =&-H^4M_F^2a(x)a(x')\tr\Big\{(\ii\sla\pd_\mu+\wt{M}_F)(A_++A_-\ga^0)(\ii\sla\pd_{\mu'}+\wt{M}_F)(A_+ +A_- \ga^0)\Big\}\n\\
 =&-H^4M_F^2a(x)a(x')\n\\
 &~\times\Big\{\tr[\ga^m\ga^n]e_m^\mu e_n^{\nu'}(\ii\pd_\mu A_+)(\ii\pd_{\nu'} A_+ )+\tr[\ga^m\ga^0\ga^n\ga^0]e_m^\mu e_n^{\nu'}(\ii\pd_\mu A_-)(\ii\pd_{\nu'} A_-)\n\\
 &+\ii \wt{M}_F\tr[\ga^m\ga^0]\Big[A_- e_m^\mu \pd_\mu A_++A_+ e_m^\mu \pd_\mu A_- +A_+e_m^\mu \pd_{\mu'}A_- +A_-e_m^{\mu'}\pd_{\mu'} A_+ \Big]\n\\
 &+\tr[\mb{1}]\wt{M}_F^2A_+A_+ +\tr[\ga^0\ga^0]\wt{M}_F^2A_-A_- \Big\}\n\\
=&-4H^4M_F^2a(x)a(x')\n\\
&~\times \Big\{
\big(e_m^\mu \pd_\mu A_+ \big)\big(e^{m\nu'} \pd_{\nu'} A_+ \big)-\big(e_m^\mu \pd_\mu A_- \big)\big(e^{m\nu'} \pd_{\nu'} A_- \big)-2H^2\tau\tau'\pd_\tau A_-\pd_\tau'A_-\n\\
 &+\ii \wt{M}_F\Big[(-H\tau)A_- \pd_\tau A_+ +(-H\tau)A_+ \pd_\tau A_- +(-H\tau')A_+ \pd_{\tau'}A_- +(-H\tau')A_- \pd_{\tau'} A_+ \Big]\n\\
 &+\wt{M}_F^2A_+^2+\wt{M}_F^2A_-^2\Big\},
\end{align}
where $\wt{M}_F\equiv M_F+2H$. The additional $2H$ is from the covariant derivative acting on the constant spinor.
We are interested in the late time limit of the two coordinates $x=(\tau,\mb x)$ and $x'=(\tau',\mb x')$ where the spatial distance $X\equiv |\mb x-\mb x'|$ is fixed while $\tau,\tau'\to 0$. In this limit, we have,
\bge
  A_\pm \sim \FR{H^2X^2}{8\pi^{5/2}}\Big(\FR{\tau\tau'}{X^2}\Big)^{2+\ii {\mu_{1/2}}}\Gamma(1+\ii {\mu_{1/2}})\Gamma(\FR{1}{2}-\ii {\mu_{1/2}})\pm({\mu_{1/2}}\to-{\mu_{1/2}}).
\ede
Plug this result into above expression we get,\begin{align}
&\Big\la\big[\ob{\Psi}\sla{\nabla}\Psi(x)\big]\big[\ob{\Psi}\sla{\nabla}\Psi(x')\big]\Big\ra\n\\
&=\FR{H^4M_F^4}{32\pi^5}\bigg[(7+24\ii)\Gamma^2(\ii {\mu_{1/2}})\Gamma^2(\FR{1}{2}-\ii {\mu_{1/2}})\Big(\FR{\tau\tau'}{X^2}\Big)^{2+2\ii {\mu_{1/2}}}+\text{c.c.}\bigg].
\end{align}
Then the three-point function from the spinor loop is,
\begin{align}
&\big\la\de\phi(\mb k_1)\de\phi(\mb k_2)\de\phi(\mb k_3)\big\ra_{\Psi4}\n\\
&=\bigg[\FR{f_{\Psi }'(X_0)}{1+f_\Psi(X_0)}\bigg]^{2}\FR{H^7M_F^2\dot\phi_0}{2\pi^4k_S^6}C_{\Psi}({\mu_{1/2}})\Big(\FR{k_L}{k_S}\Big)^{-1+2\ii {\mu_{1/2}}}+\text{c.c.},
\end{align}
where the mass dependent coefficient $C_{\Psi 4}({\mu_{1/2}})$ is given by,
\begin{align}
  C_{\Psi}({\mu_{1/2}})
  =&~ \mu^4_{1/2} (-24+7\ii)(1+\ii {\mu_{1/2}})(3+2\ii {\mu_{1/2}})
 \cosh^3(\pi {\mu_{1/2}})\sinh(\pi {\mu_{1/2}})\n\\
 &~\times\Gamma^2(\FR{1}{2}-\ii {\mu_{1/2}})\Gamma(-2-4\ii {\mu_{1/2}})\Gamma^2(\ii {\mu_{1/2}})\Gamma^2(2\ii {\mu_{1/2}}).
\end{align}
Note that this amplitude does not have a pole.
A special feature of fermion loop is that the exponent of momentum ratio always contains an imaginary part $2\ii {\mu_{1/2}}$ so long as the fermion is massive, and therefore, the signal of massive fermion loop is always oscillatory.

\subsection{$F_{\mu\nu}F^{\mu\nu}$}

Finally, we consider the gauge boson loop. Just like the case of Higgs field, we can consider one real component of vector bosons with mass $M_A$, and the result apply equally to $W/Z$ and photon, with mass and degrees of freedom properly adjusted.

We denote the vector boson field being considered as $A_\mu$ and the quadratic part of its kinetic term as $F_{\mu\nu}F^{\mu\nu}$. If $A_\mu$ is a component of non-Abelian gauge field ($W$ or $Z$), then it is understood that the self-interaction part is excluded from $F_{\mu\nu}$ since it does not contribute at 1-loop level. Then the operator $\O_\al$ we are interested in can be written in terms of properly normalized field strength $F_{\mu\nu}$ as,
\bge
  S\supset-\FR{1}{4\big(1+f_A(X_0)\big)}\int\di^4x\sqrt{-g}f_{A}(X)F_{\mu\nu}F^{\mu\nu}.
\ede
Then we need to evaluate the following 2-point correlation function,
\begin{align}
\Big\la\big[F_{\mu\nu}F^{\mu\nu}(x)\big]\big[F_{\rh\si}F^{\rh\si}(x')\big]\Big\ra=\Big[\big(\nabla_\mu\nabla_{\rh'} G_{\nu\si'}(x,x')-(\mu\leftrightarrow\nu)\big)-(\rh'\leftrightarrow\si')\Big]^2,
\end{align}
in which the vector propagator is given by (\ref{VP}) and the covariant derivatives can be taken using the formulae in Appendix \ref{AppEdS}. As before, the result can be expanded at late time limit $\tau,\tau'\to 0$ as,
\begin{align}
&\Big\la\big[F_{\mu\nu}F^{\mu\nu}(x)\big]\big[F_{\rh\si}F^{\rh\si}(x')\big]\Big\ra\n\\
&= \FR{27H^{12}}{16\pi^5M_A^4}\bigg[(23-6\mu_1)^2\Gamma^2(\mu_1)\Gamma^2(\FR{5}{2}-\mu_1)\Big(\FR{\tau\tau'}{X^2}\Big)^{3-2\mu_1}+(\mu_1\to-\mu_1)\bigg],
\end{align}
where $\mu_1=\sqrt{1/4-(M_A/H)^2}$. Then the three-point function in squeezed limit from this loop is,
\begin{align}
&\big\la\de\phi(\mb k_1)\de\phi(\mb k_2)\de\phi(\mb k_3)\big\ra_{F4}\n\\
&=\bigg[\FR{f_{A}' (X_0)}{1+f_A(X_0)}\bigg]^2\FR{27H^{13}\dot\phi_0}{\pi^4M_A^4(2k_S)^6} C_{A}(\mu_1)\Big(\FR{k_L}{2k_S}\Big)^{-2\mu_1}+(\mu_1\to-\mu_1),
\end{align}
where the mass-dependent coefficient $C_{A}(\mu_1)$ is,
\begin{align}
C_{A}(\mu_1)\equiv&-(2-\mu_1)(3-2\mu_1)(23-6\mu_1)^2\n\\
&\times\Gamma^2(\mu_1)\Gamma(-4+4\mu_1)\Gamma^2(\FR{5}{2}-\mu_1)\Gamma^2(2-2\mu_1)\sin^3(\pi\mu_1)\cos(\pi\mu_1).
\end{align}
This amplitude has an unphysical pole at $\mu_1=1/4$.

\subsection{Summary of the correlation functions}

In above calculations we have obtained the squeezed limit of three-point function of inflaton fluctuations, contributed by SM loops. Now let us summarize the results more systematically. We shall express these results in terms of curvature fluctuations $\zeta=-H\delta\phi/\dot\phi_0$. According to the standard parameterization of three-point function of $\zeta$ \cite{Chen:2010xka, Wang:2013eqj},
\begin{eqnarray}
\label{zeta3para}
\langle \zeta(\mb k_1)\zeta(\mb k_2)\zeta(\mb k_3)\rangle \equiv
S(k_1,k_2,k_3) \frac{1}{(k_1k_2k_3)^2}
P_{\zeta}^2 (2\pi)^7
\delta^3(\mb k_1+\mb k_2+\mb k_3),
\end{eqnarray}
where $P_\zeta \equiv H^2/(8\pi^2 \Mp^2 \epsilon)$ is the power spectrum of the curvature perturbation.
Now we take the squeezed limit $k_{1,2}\gg k_3$ and using the notation $k_S\equiv k_1\simeq k_2$ and $k_L\equiv k_3$, we can define the magnitude of non-Gaussianity $f_{NL}$ as,
\begin{eqnarray}
S(k_L,k_S) \sim f_{NL} \left( \frac{k_{L}}{k_{S}} \right)^\ga ~,
\end{eqnarray}
up to a normalization numerical factor. Below we collect the squeezed limit shape functions $S$ for bispectra involving various SM operators $\O_\al$ calculated above,
\begin{align}
\label{SH2}
S_{H}=&~ \left [ \frac{f_H'(X_0)}{1+f_{DH}(X_0)}  \right ]^2
\FR{\dot\phi_0^2}{2\pi^4}\bigg[C_{H}(\mu_h)\Big(\FR{k_L}{2k_S}\Big)^{2-2\mu_h} +(\mu_h\to-\mu_h)\bigg],\\
\label{SH4}
S_{DH}=&~ \left [ \frac{f_{DH}'(X_0)}{1+f_{DH}(X_0)}  \right ]^2
\FR{H^4\dot\phi_0^2}{8\pi^4}\bigg[C_{DH}(\mu_h)\Big(\FR{k_L}{2k_S}\Big)^{2-2\mu_h} + (\mu_h\to-\mu_h) \bigg],\\
\label{SPsi4}
S_{\Psi}=&~ \left [ \frac{f_{\Psi}'(X_0)}{1+f_{\Psi}(X_0)}  \right ]^2
\FR{H^4\dot\phi_0^2  \mu_{1/2}^2}{2\pi^4}\bigg[C_{\Psi}({\mu_{1/2}})\Big(\FR{k_L}{k_S}\Big)^{1+2\ii {\mu_{1/2}}}+\text{c.c.}\bigg],\\
\label{SF4}
S_{A}=&~ \left [ \frac{f_{A}'(X_0)}{1+f_{A}(X_0)}  \right ]^2
\FR{27H^{8}\dot\phi_0^2}{16\pi^4 M_A^4}\bigg[C_{A}(\mu_1)\Big(\FR{k_L}{2k_S}\Big)^{2-2\mu_1}+(\mu_1\to-\mu_1)\bigg] ~,
\end{align}
where $\mu_h=\sqrt{9/4-(M_H/H)^2}$, ${\mu_{1/2}}\equiv M_F/H$ and
$\mu_1=\sqrt{1/4-(M_A/H)^2}$. As we can see, the exponent $\ga$ can be either real or complex, depending on the mass of SM fields in the loop, and we would see characteristic power-law behavior or oscillatory behavior, respectively.
In particular, the oscillatory dependence on the momentum ratio directly encodes the scale factor evolution of the inflationary background \cite{Chen:2015lza}, because the massive fields can be regarded as primordial standard clocks. If detected, these signals would provide a direction evidence for inflation.

To be observable, the SM masses need to be around the Hubble scale or less, as too large mass would suffer from strong Boltzmann suppression.
For these signals, if $f_{NL} > 0.01$, we can hope to see them or even distinguish different $\ga$'s in the future 21cm surveys \cite{Meerburg:2016zdz} or large scale structure surveys \cite{Sefusatti:2012ye,Norena:2012yi,Gleyzes:2016tdh}, at least in principle.

\subsection{Observational consequences}


In this subsection, we discuss a few interesting features of the amplitudes of the bispectra obtained in this section.

Firstly, to have an estimate for the magnitude of non-Gaussianity with the $C$-coefficients, we note that the size of those $C$-coefficients can be approximated with
\begin{align}
  |C_H(\mu_h)| \sim &~ \frac{1.5}{\cosh(2\pi|\mu_h|)}~,
  \\
  |C_{DH}(\mu_h)| \sim &~ \frac{30}{\cosh(2\pi|\mu_h|)}~,
  \\
  |C_{\Psi}(\mu_{1/2})| \sim &~ \frac{23}{\cosh(2\pi|\mu_{1/2}|)}~,
  \\
  |C_{A}(\mu_1)| \sim &~\frac{1800}{\cosh(2\pi|\mu_1|)}~,
\end{align}
when $\mu_h$ and $\mu_1$ are purely imaginary and $\mu_{1/2}$ is real. This approximation works quite well before the $C$-coefficients are exponentially suppressed. When $|\mu| \gg 1$, the approximation underestimates the $C$-coefficients, but is a good approximation on logarithmic scale.

Therefore, as an order-of-magnitude estimate, we have, for example,
\begin{align}
   f_{NL}(F^2) \sim \frac{30\dot\phi_0^2 H^8}{M_A^4 \cosh(2\pi|\mu_1|)} \left [ \frac{f_{A}'(X_0)}{1+f_{A}(X_0)}  \right ]^2 ~,
\end{align}
as well as similar expressions for the other fields.

For mass of SM fields of order Hubble, to make $f_{NL}\sim 1$, it is required that $f_{A}'^{\,2}(X_0)\sim H^{-4}\dot\phi_0^{-2}$ (similarly, $f_{H}'^{\,2}(X_0)\sim \dot\phi_0^{-2}$ and $f_{DH}'^{\,2}(X_0)\sim H^{-4}\dot\phi_0^{-2}$), which should be in principle attainable even if the smallness of $f_{\al}(X_0)$ is assumed. This parameter region is not likely to be natural, but it would be good to find realistic and complete inflation models so that this parameter region is realized. On the other hand, if one takes the simplest choice $f_\al(X)\propto X$, then it is easy to see that the smallness of $f_\al(X)$ and the observability of the oscillatory/power-law signal cannot be both satisfied. In this case, we would either see a rather arbitrary mass spectrum of the SM background or very clean background without any detectable SM signals at all.

On the other hand, the calculation in this section can be readily extended to new particles in beyond SM new physics. In particular, when those new particles are gauge singlet so that they can be produced singly, they can contribute to the squeezed limit of bispectrum at tree level. It is expected that the signal of such tree diagrams can be much more significant than the SM signals even if the couplings between new particle and inflaton are of the simplest type, i.e. $f(X)\propto X$. Examples of such particles include various type of axions in string theory or Peccei-Quinn type theories, right-handed sneutrino in SUSY theories. With some luck, the cosmological collider may be a good discovery machine for these beyond SM new particles.

Secondly, note that the masses of the SM fields depend on $f(X_0)$, and the non-Gaussianities depend on $f'(X_0)$, a consistency relation can be constructed by making use of the scale dependence of mass parameters. Taking the gauge bosons for example, one can calculate the Weinberg angle from the gauge boson's mass ratio as,
\begin{align}
  \frac{M_Z^2}{M_W^2}-1=\tan^2\theta_W  =\frac{1+f_W}{1+f_B}\tan^2\theta_W^{\text{SM}}~.
\end{align}
Neglecting the running of the SM Weinberg angle $\theta_{W}^{\text{SM}}$ for simplicity, the scale dependence of $\tan^2\theta_W$ gives
\begin{align}
  \frac{d \ln \tan^2\theta_W}{d\ln k} =   \frac{d\ln(1+f_W)}{d\ln k} - \frac{d\ln(1+f_B)}{d\ln k}
  = \left (  \frac{f_W'}{1+f_W} - \frac{f_B'}{1+f_B}  \right ) \frac{\dot\phi_0\ddot\phi_0}{H} ~.
\end{align}
This relation can be readily related to Eq.~\eqref{SF4}. If we normalize the non-Gaussianity of Eq.~\eqref{SF4} as,
\begin{align}
  &f_{NL}^W \equiv
  N_W \left ( \frac{f_W'}{1+f_W}  \right )^2 \frac{27H^8\dot\phi_0^2}{16\pi^4M_W^4} |C_A(\mu_W)| ~,\\
  &f_{NL}^Z \equiv
  N_Z\left(  \frac{f_W'}{1+f_W}\cos^2\theta_W+\frac{f_B'}{1+f_B}\sin^2\theta_W \right)^2 \frac{27H^8\dot\phi_0^2}{16\pi^4M_Z^4} |C_A(\mu_Z)| ~,
\end{align}
where $N_W=2$ and $N_Z=1$ are from counting of field content. We have,
\begin{align}
   \frac{d \ln \tan^2\theta_W}{d\ln k} = \frac{\pi (\eta - 2\epsilon) }{3\sqrt{3 P_\zeta}\sin^2\theta_W}
   \left [
      \frac{M_W^2}{H^2}\sqrt{\frac{f_{NL}^W}{N_W |C_A(\mu_W)|} }
    - \frac{M_Z^2}{H^2}\sqrt{\frac{f_{NL}^Z}{N_Z |C_A(\mu_Z)|} }
   \right ]~,
\end{align}
where $\epsilon\equiv -\dot H/H^2$ and $\eta\equiv \dot\epsilon/(H\epsilon)$ are slow roll parameters. This equation can further be recasted using the spectral index of curvature perturbation $n_s$ and the tensor-to-scalar ratio $r$ as,
\begin{align}\label{eq:cons-rel}
   \frac{d \ln \tan^2\theta_W}{d\ln k} = \frac{\pi (1-n_s - \frac{1}{4} r) }{3\sqrt{3 P_\zeta}\sin^2\theta_W}
   \left [
      \frac{M_W^2}{H^2}\sqrt{\frac{f_{NL}^W}{N_W |C_A(\mu_W)|} }
     -\frac{M_Z^2}{H^2}\sqrt{\frac{f_{NL}^Z}{N_Z |C_A(\mu_Z)|} }
   \right ]~,
\end{align}
In the near future, one shall be able to determine $1-n_s-\frac{1}{4} r$ very accurately. To verify \eqref{eq:cons-rel}, very precise measurements of the mass parameters $M_W$ and $M_Z$ are needed. This would be a very challenging test for the future non-Gaussianity measurement.

Thirdly, so far we have not assumed any relations between the $f(X)$ parameters. In specitific inflation models, there may be various relations between $f(X)$ couplings for different SM fields. If it is true, additional predictions can be made. For example, if the inflaton is coupled to the SM sector via a common coupling $\mathcal{L} = f(X)\mathcal{L}_{\text{SM}}$, then we have $\theta_W = \theta_W^{\text{SM}}$, along with some other predictions. It is interesting to study the implications of those assumptions in the model building point of view.

\section{SM Fingerprints of Higgs Inflation}
\label{Sec_Higgs}

In previous sections we have focused on non-Higgs inflation models, assuming that Higgs field does not develop nonzero VEV during inflation and that the inflaton couples to SM fields through derivative coupling due to approximate shift symmetry. Both of two assumptions are not valid if the inflaton is just the SM Higgs boson itself. Given the fact that the Higgs boson is the only fundamental scalar particle in SM, and also the only fundamental scalar particle experimentally discovered, it is both natural and important to study the possibility that the Higgs field itself is the inflaton. In fact, the Higgs boson can indeed be the inflaton, and the simplest model built on this assumption is consistent with basically all known results from both particle experiments and cosmological observations, with one important caveat about the quantum correction of the Higgs potential at high energies which we shall comment on below. This is the model firstly proposed by \cite{Bezrukov:2007ep} and we shall refer to it as the original Higgs inflation model.

A potential problem for original Higgs inflation is the Higgs instability mentioned before. Higgs self-coupling may decrease to negative values when the energy scale is larger than $10^{11}$GeV, due to the renormalization group running, and this scale is much smaller than the scale of pre-normalized Higgs field, which is typically $10^{16}$GeV during inflation\footnote{Here the pre-normalized Higgs field refers to the Higgs field in Einstein frame without canonical normalization, given in (\ref{h2}) below. During inflation there are several relevant scales, including the magnitude of inflaton, the magnitude of energy density, the magnitude of Hubble scale, and one must be careful when making comparisons. When considering renormalization group running of Higgs potential, it is the pre-normalized Higgs field that should be compared with the renormalization scale.}. Depending on the uncertainties in the measured value of top quark mass, the sign-changing scale of Higgs potential may be much higher than $10^{11}$GeV, since this scale is exponentially sensitive to the mass input of Higgs boson and top quark. But even after taking this into account, it would still be some tension between the positiveness of Higgs potential during Higgs inflation and the Higgs and top mass measurements. May or may not be a fatal problem for the original Higgs inflation \cite{Bezrukov:2014ipa}, the Higgs instability has nevertheless motivated a lot of studies on the possible extension of original Higgs inflation to various new physics scenario, and by far it is clear that Higgs inflation can be realized in many different models, including simple extensions of SM and more complete new physics models such as supergravity GUTs \cite{He:2014ora,Xianyu:2014eba,Ellis:2014dxa,Ge:2016xcq,Ellis:2016spb,Hamada:2014wna,Hamada:2014xka,Rubio:2014wta}.

Given the abundance of Higgs inflation models, a natural question to ask is, can we find any universal feature of these models that can separate them apart from other non-Higgs inflation theories. At the level of linear perturbation theory, this is almost impossible due to the lack of observables. In fact, the original Higgs inflation has almost identical predictions to power spectrum of both scalar mode and tensor mode with any single field slow roll model with a exponentially flat potential, of which the Starobinsky model is a notable example. If we consider generalization of original Higgs inflation to other new physics theories, one can even achieve more wider range of predictions, with tensor-to-scalar ratio $r$ varying from $\order{0.1}$ to $\order{10^{-7}}$.

Therefore, we need to go beyond linear perturbations. Here a very important clue for distinguishing Higgs inflations from other inflation models is that the electroweak symmetry is spontaneously broken in the former case. Indeed, the Higgs inflaton during inflation generally acquires an extremely large background value, and the excursion of the canonically normalized Higgs field VEV can be at the same order with the Planck scale.\footnote{This huge VEV of Higgs field does not violate perturbativity of the model, as has been carefully studied in \cite{Ren:2014sya}.} An immediate consequence of this observation is that the SM mass spectrum will be vastly different from the spectrum in non-Higgs inflation models. Therefore, in the case of Higgs field, all SM fields that receive masses from Higgs VEV can acquire huge mass during inflation due to the huge background value of Higgs field.

In original Higgs inflation, both the SM spectrum and the couplings between Higgs inflaton and other SM fields can be unambiguously determined, and therefore, one can in principle make a rather definite prediction on these signals. Given the unique feature of spontaneous electroweak symmetry breaking, this signals can be a distinctive feature of Higgs inflation, and for this reason we call it the ``SM fingerprints'' of the Higgs inflation.

In this section we shall carry out an analysis of Higgs inflation parallel to previous two sections for non-Higgs inflation. We shall review Higgs inflation very briefly, and then work out the corresponding SM spectrum, as well as their signals in the squeezed limit of bispectrum.

\subsection{Higgs Inflation and the SM Spectrum}

The original Higgs inflation makes use of the SM Higgs potential, plus the crucial ingredient of non-minimal coupling between Higgs field and Ricci scalar. The inflation scale is much higher than the electroweak scale, so that the negative quadratic term in the Higgs potential can be safely neglected when studying physics during inflation\footnote{But the negative quadratic term is crucial in an implicitly way, because it provides the mass to Higgs boson at the electroweak scale, and the Higgs mass, as as a input for renormalization running.}. With this point in mind, we can write down the SM action in a general curved background as follows,
\begin{align}
\label{SJ}
  S_J=&~\int\di^4x\,\sqrt{-g}\bigg[\Big(\FR{1}{2}M^2+\xi\mb H^\dag \mb H\Big)R-|\D_\mu\mb H|^2-\lam(\mb H^\dag \mb H)^2-\FR{1}{4}\sum_{I}F_{I\mu\nu }^a F_I^{\mu\nu a}\n\\
    &~+\sum_i \ii\ob\Psi_i \sla{\mathscr{D}}\Psi_i-(y_\ell \ob{L}\mb H\ell_R+y_d\ob{Q}\mb H d_R+y_u\ob{Q}\wt{\mb H} u_R+\text{c.c.})\bigg],
\end{align}
in which we have the Higgs doublet $\mb H$ coupled to the Ricci scalar $R$ with nonminimal coupling $\xi$. We denote all gauge field strength of the SM gauge group collectively as $F_{I\mu\nu}^a$, and denote all SM fermions collectively as $\Psi_i$. In Yukawa terms, we have the left-handed lepton doublets $L$ and quark $Q$, together with right handed singlets $\ell_R$, $u_R$, and $d_R$. The covariant derivative $\D_\mu\mb H$ is associated with the SM gauge group under which the Higgs doublet is charged, while the covariant derivative $\mathscr{D}\Psi_i$ contains both SM gauge fields and spin connection. The action (\ref{SJ}) is conventionally said to be written in the Jordan frame. The nonminimal coupling term in this action can be eliminated by a ``frame transformation'', i.e. the following field redefinition,
\bge
\begin{aligned}
  &g_{\mu\nu} \to \Omega^{-2}g_{\mu\nu},
  &&\Omega^2\equiv \FR{M^2+2\xi\mb H^\dag\mb H}{\Mp^2},
\end{aligned}
\ede
and we can take $M=\Mp$ for simplicity. As a result, we reach the following action which is said to be written in the Einstein frame \cite{Xianyu:2013rya,Ren:2014sya},
\begin{align}
\label{SE}
  S_E=&~\int\di^4x\,\sqrt{-g}\bigg[\FR{1}{2}\Mp^2 R-\FR{1}{\Omega^2}|\D_\mu\mb H|^2-\FR{\lam}{\Omega^4}(\mb H^\dag \mb H)^2-\FR{3\xi^2}{\Mp^2\Omega^4}\big(\nabla_\mu(\mb H^\dag \mb H)\big)^2\n\\
  &~-\FR{1}{4}\sum_{I}F_{I\mu\nu}^a F_I^{\mu\nu a}+\sum_i \ii\ob\Psi_i \sla{\mathscr{D}}\Psi_i-\FR{1}{\Omega}(y_\ell \ob{L}\mb H\ell_R+y_d\ob{Q}\mb H d_R+y_u\ob{Q}\wt{\mb H} u_R+\text{c.c.})\bigg],
\end{align}
where we have also redefine the fermion fields according to $\Psi_i\to \Omega^{3/2}\Psi_i$ so that their kinetic terms are canonically normalized. Compared with Jordan frame action (\ref{SJ}), the Einstein frame action (\ref{SE}) does not contain nonminimal coupling, but instead, there appear several new features: 1) The Higgs field is no longer canonically normalized (and therefore, a further normalization of Higgs field is needed); 2) Higgs field receives new higher-dimensional derivative couplings; 3) The Higgs potential is rescaled by a factor of $\Omega^{-4}$ and similarly the Yukawa terms by a factor of $\Omega^{-1}$.

In our present study it is convenient to work in the unitary gauge in which the Higgs doublet can be parameterized by a single real component $h$ via $\mb H=(0,h/\sqrt{2})^T$. By examining the kinetic term of $h$, we can find the corresponding normalized field $\phi$, which is related to $h$ via,
\bge
  \FR{\di\phi}{\di h}=\FR{\sqrt{\Omega^2+6\xi^2h^2/\Mp^2}}{\Omega^2}.
\ede
In original Higgs inflation, we have $h\gg \Mp/\xi$ during inflation era, and therefore, we can make the following approximation,
\bge
\label{h2}
  h^2(\phi)\simeq \FR{\Mp^2}{\xi}\Big(e^{\sqrt{2/3}\phi/\Mp}-1\Big),~~~~~~\Omega^2(\phi)\simeq e^{\sqrt{2/3}\phi/\Mp}.
\ede

Now we figure out the SM spectrum during inflation. Firstly, the Higgs field has only one physical component which we have identified to be the inflaton. As in any slow-roll model, the Higgs inflaton is nearly massless during inflation, and its self-interaction is extremely weak. On the other hand, due to the presence of a huge Higgs VEV, $W/Z$ boson and all charged fermions are massive during inflation, while the photon and the gluon remain massless. Meanwhile, both $W/Z$ and charged fermions interact with Higgs inflaton directly, i.e. no spacetime derivative involved at the leading order. The masses of $W/Z$ and charged fermions and their interactions with Higgs inflaton can be easily found by examining  the part of the Lagrangian which is quadratic in these fields. Firstly, for the $W/Z$ boson,
\begin{align}
&-\FR{g^2h^2}{4\Omega^2}\big(W_\mu^+W^{-\mu}+\FR{1}{2\cos^2\theta_W}Z_\mu Z^\mu\big)\n\\
&=-\FR{g^2}{4}\big(W_\mu^+W^{-\mu}+\FR{1}{2\cos^2\theta_W}Z_\mu Z^\mu\big)\Big[\mathcal{V}^2+\mathcal{G}_1\de\phi+\FR{1}{2}\mathcal{G}_2\de\phi^2+\cdots\Big],
\end{align}
in which we have separated in the Higgs inflaton $\phi$ into the background value and the fluctuation, $\phi=\phi_0+\de\phi$, and expanded the quantity $h^2/\Omega^2$ in terms of $\de\phi^2$ up to quadratic order, which is all we need. The various effective couplings in above expression are defined as follows,
\begin{align}
  \mathcal{V}=&~\FR{h_0}{\Omega_0},\\
 \mathcal{G}_1=&~\FR{\di}{\di\phi}\FR{h^2}{\Omega^2}\bigg|_{\phi=\phi_0}=\sqrt{\FR{2}{3}}\FR{\Mp}{\xi\Omega_0^2},\\
 \mathcal{G}_2=&~\FR{\di^2}{\di\phi^2}\FR{h^2}{\Omega^2}\bigg|_{\phi=\phi_0}=-\FR{2}{3\xi\Omega_0^2},
\end{align}
where the subscript $0$ indicates that the quantity is to be evaluated at $\phi=\phi_0$. In the same way we can also find the quadratic part of fermionic action as follows,
\begin{align}
-\FR{y_i}{\sqrt 2}\FR{h}{\Omega}\ob{\Psi}_i\Psi_i
=&-\FR{y_i}{\sqrt 2}\ob{\Psi}_i\Psi_i\Big[\mathcal{V}+\mathcal{F}_1\de\phi+\FR{1}{2}\mathcal{F}_2\de\phi^2+\cdots\Big],
\end{align}
where the two new couplings are defined as,
\begin{align}
 \mathcal{F}_1=&~\FR{\di}{\di\phi}\FR{h}{\Omega}\bigg|_{\phi=\phi_0}=\FR{\Mp}{\sqrt 6\xi h_0\Omega_0},\\
 \mathcal{F}_2=&~\FR{\di^2}{\di\phi^2}\FR{h}{\Omega}\bigg|_{\phi=\phi_0}=-\FR{\Mp^2(2\Omega_0^2-1)}{6\xi^2\Omega_0h_0^3}.
\end{align}

An interesting feature of above results is that the SM spectrum during Higgs inflation is very different from the non-Higgs inflation models, but qualitatively similar to the SM spectrum in electroweak broken phase, with the Higgs VEV $v\simeq 246$GeV replaced by $\mathcal{V}$. And also, the quantity $\mathcal{V}$ is actually not a constant during inflation and therefore, the masses of all SM particles during Higgs inflation are changing as the Higgs inflaton rolls down along its potential. However, this changing is significant only for the final dozens of $e$-folds during observable inflation, which are very difficult to observe, so we shall treat the mass as constant, taking the value $h_0/\Omega_0$ at the onset of observable inflation.

\subsection{Signals in Bispectrum}

In this subsection we compute the squeezed limit of relevant bispectrum of inflaton perturbations to show the shape and strength of the SM fingerprints in Higgs inflation. The amplitude we are going to calculate is still (\ref{3ptInt}), and this amplitude can again be related to corresponding four-point amplitude of conformal scalars (\ref{4ptInt}). The difference from the previous section is that the SM fields couple to Higgs inflaton with non-derivative interaction, while in previous section we only considered derivative coupling. Therefore, in previous section, the 3-point function of inflaton fluctuations is obtained from the conformal amplitude $\I_c(k_{12},k_{34},k_I)$ by acting differential operator, but in the current situation, we have an integral operator instead. That is,
\begin{align}
\I(k_1,k_2,k_3)=\big(\mathcal{K}_{12}^2+k_{12}\mathcal{K}_{12}+k_1k_2\big)\big(\mathcal{K}_3^2+k_3\mathcal{K}_3\big)\I_c(k_{12},k_3,k_3),
\end{align}
where $\mathcal{K}_{12}$ is an integral operator defined as follows,
\bge
  \mathcal{K}_{12}f(k_{12})=\int_{k_{12}}^\infty \di k_{12}'\,f(k_{12}'),
\ede
and $\mathcal{K}_3$ is defined similarly.

With this slight modification, it is straightforward to work out the non-local part of the 2-point function of SM operators at one-loop level. For gauge fields, we have,
\begin{align}
&\la A^2(x)A^2(x')\ra\n\\
&=\FR{9H^8}{8\pi^5m_A^4}\bigg[(4-\mu_1)\Gamma^2(\mu_1)\Gamma^2(\FR{5}{2}-\mu_1)\Big(\FR{\tau\tau'}{X^2}\Big)^{3-2\mu_1}+(\mu_1\to-\mu_1)\bigg],
\end{align}
in which $\mu_1=\sqrt{1/2-(m_A/H)^2}$ and for spinor fields, we have,
\begin{align}
&\la \ob\Psi\Psi(x)\ob\Psi\Psi(x')\ra\n\\
&=\FR{H^4m_F^2}{32\pi^5}\bigg[(7+24\ii)\Gamma^2(\ii {\mu_{1/2}})\Gamma^2(\FR{1}{2}-\ii {\mu_{1/2}})\Big(\FR{\tau\tau'}{X^2}\Big)^{2+2\ii {\mu_{1/2}}}+\text{c.c.}\bigg],
\end{align}
in which ${\mu_{1/2}}=m_F/H$. Then through the procedure parallel with the last section, we find the squeezed limit of 3-point function for inflaton fluctuations contributed from $\la A^2A^2\ra$ to be,
\begin{align}
\la\de\phi(\mb k_1)\de\phi(\mb k_2)\de\phi(\mb k_3)\ra_{A}=\FR{9\mathcal{G}_1\mathcal{G}_2H^6}{32\pi^4m_A^4k_S^6}\wt{C}_{A}(\mu_1)\Big(\FR{k_L}{2k_S}\Big)^{-2\mu_1}+(\mu_1\to-\mu_1),
\end{align}
where
\begin{align}
\wt{C}_A(\mu_1)=&~\mu_1(2-\mu_1)(4-\mu_1)(1-2\mu_1)(3-2\mu_1)(2+3\mu_1+2\mu_1^2)\n\\
&~\times\Gamma(-3+4\mu_1)\Gamma^2(-1+\mu_1)\Gamma^2(-2\mu_1)\Gamma^2(\FR{3}{2}-\mu_1)\n\\
&~\times\sin^2(\pi\mu_1)\sin(2\pi\mu_1).
\end{align}
For spinor fields, we have,
\begin{align}
\la\de\phi(\mb k_1)\de\phi(\mb k_2)\de\phi(\mb k_3)\ra_{\Psi}=\FR{y^2\mathcal{F}_1\mathcal{F}_2H^4}{2\pi^4k_S^6}\wt{C}_\Psi({\mu_{1/2}})\Big(\FR{k_L}{k_S}\Big)^{-1+2\ii {\mu_{1/2}}}+\text{c.c.},
\end{align}
where
\begin{align}
\wt{C}_\Psi({\mu_{1/2}})=&~\FR{24+7\ii}{1-\ii {\mu_{1/2}}+2u^2}\mu_{1/2}^3(3\ii-2{\mu_{1/2}})(-4+5\ii {\mu_{1/2}}+2\mu_{1/2}^2)\n\\
&~\times\Gamma^2(-\FR{1}{2}-\ii {\mu_{1/2}})\Gamma(-1-4\ii {\mu_{1/2}})\Gamma^2(\ii {\mu_{1/2}})\Gamma^2(2\ii {\mu_{1/2}})\n\\
&~\times\cosh^3(\pi {\mu_{1/2}})\sinh(\pi {\mu_{1/2}}).
\end{align}
We then find the squeezed limit bispectrum using the definition of (\ref{zeta3para}),
\begin{align}
S_A=&~\FR{9\mathcal{G}_1\mathcal{G}_2H\dot\phi_0}{8\pi^4m_A^4}\wt{C}_A(\mu_1)\Big(\FR{k_L}{2k_S}\Big)^{2-2\mu_1}+(\mu_1\to-\mu_1),\\
S_\Psi=&~\FR{\mathcal{F}_1\mathcal{F}_2\dot\phi_0}{\pi^4H}\wt{C}_\Psi({\mu_{1/2}})\Big(\FR{k_L}{k_S}\Big)^{1+2\ii {\mu_{1/2}}}+\text{c.c.}.
\end{align}
It is however unfortunate that the ``fingerprints'' of SM fields in original Higgs inflation is far too weak to be observable. The physical reason is clear: the Boltzmann suppression is severe for very heavy fields like top quark and $W/Z$ bosons, while the coupling to Higgs inflaton is too small for light fields like charged fermions of first generation. However, it would be interesting to seek variations of original Higss inflation model so that the SM fingerprints can be observable, and the calculation presented in this section would be helpful in those cases.

\section{Discussions}
\label{Sec5}

In this paper we have studied the Standard Model spectrum during the inflationary era of our universe, assuming a generic single field slow-roll model of inflation. The spectrum turns out to be quite different from both the ordinary SM mass spectrum in the electroweak broken phase and the trivial massless spectrum in electroweak symmetric phase.
Notably, the masses of many SM fields, such as the Higgs and some gauge bosons, are lifted to be around the Hubble scale $H$ due to quantum corrections.
The details of this spectrum depends on the background rolling velocity of inflaton $\dot\phi_0$, the effective couplings between SM fields and inflaton, as well as on the quantum corrections. Manipulations in Euclidean de Sitter space have played a crucial role and have brought great simplification to our calculation.

For non-Higgs inflation, depending on the strength of the interactions between SM fields and inflaton, the SM spectrum can be very different in different inflation models. For example, if the interactions $f_\al(X)\mathcal{O}_\al$ satisfy $f_\al(X_0)\ll 1$, then the SM spectrum turns out to be quite universal and predictable, which depends only on the Hubble scale $H$ and the non-minimal coupling $\xi$ between Higgs field and Ricci scalar. However, if $f_\al(X_0)$'s are large enough, the masses of the Higgs and W/Z's can become somewhat arbitrary although the photon still remains massless.

The SM spectrum can manifest itself through the squeezed limit of bispectrum of the curvature perturbation. 
In the current work we have assumed that the inflaton is a SM singlet, so the leading order effects of SM fields are at 1-loop order.
We have computed the amplitudes and shapes of the squeezed-limit bispectra that correspond to the SM particle spectrum.
The shapes of the bispectra are determined by the mass and spin of the SM particles in the inflation background. The amplitudes are also very model-dependent. In the simplest case where the coupling $f_\al(X) \ll 1$ and $f_\al(X) \propto X$, the amplitudes are too small to be observed, so we expect no SM background for the cosmological collider. Our explicit formulae also point out the parameter space where such signals are observable.


We have adopted the effective field theory approach in this work, assuming SM coupled to a single field inflation sector, with nothing else. In particular, we do not address the naturalness problem in this work, which should be considered when new physics beyond SM is included. It would be interesting to work out more concrete examples of our calculation in various specific inflation models, in particular in those models where the inflaton-SM couplings are strong. Meanwhile, it would also be interesting to consider mass spectrum of corresponding non-Gaussian signals for new physics beyond SM, such as the right-handed neutrinos, grand unification theories, etc.

Another important direction worth exploring is the higher-spin interactions between inflaton and SM fields, which should be straightforward to work out in our current framework. The nonzero spin of such interactions can leave characteristic angular dependence in the squeezed limit of bispectrum, and it remains to be seen how can one distinguish such signal from higher-spin interactions from a genuine higher spin particle.

Finally, it is also desirable to generalize the analysis to more general effective theories of inflation models beyond slow-roll, and it remains to be seen if we would find anything dramatically new about the SM signals.

\paragraph{Acknowledgments.} We would like to thank Qing-Guo Huang, Junyu Liu, Mohammad Hossein Namjoo, Andy Strominger, and Siyi Zhou for helpful discussions. XC is supported in
part by the NSF grant PHY-1417421. YW is supported by grants HKUST4/CRF/13G and ECS 26300316 issued by the Research Grants Council (RGC) of Hong Kong. ZZX is supported in part by Center of Mathematical Sciences and Applications, Harvard University.

\begin{appendix}

\section{Euclidean dS Toolbox}
\label{AppEdS}

In this appendix we collect some basics of manipulations in $D$-dimensional Euclidean dS, which are useful in the main text.

\subsection{Preliminaries}
\label{AppEdS1}

The $D$-dimensional de Sitter space $dS_D$ can be realized as a hypersurface in $(D+1)$-dimensional Minkowski spacetime of signature $(1,D)$. With Minkowski coordinates $X^M~(M=0,1,\cdots,D)$, the hypersurface is given by the equation $-(X^0)^2+(X^1)^2+\cdots+(X^D)^2=H^{-2}$. Now if we Wick rotate the Minkowski space into Euclidean space, then $dS_D$ will be rotated into $D$-sphere $S^D$.

There are many ways of parameterizing $dS_D$ \cite{Sprad2001,Annin2012}, of which  we find two very useful coordinates of $dS_D$ in current study. One is the global coordinates. In $dS_D$ they are given by,
\bge
\begin{aligned}
  &X^0=\sinh (HT), &&X^i=\xi^i\cosh(HT), &&(i=1,\cdots, D),
\end{aligned}
\ede
where $\xi^i$'s are further parameterized by the standard spherical coordinates on the unit sphere $S^{D-1}$. The metric on $dS_D$ corresponding to the global coordinates reads,
\bge
  \di s^2=-\di T^2+(\cosh^2 T)\di\Omega_{D-1}^2.
\ede
After Wick rotation in $T$ direction, the metric above becomes the standard metric $\di\Omega_D^2$ on $S^D$ written in spherical coordinates.

The other useful coordinates $(t,x^i)$ are the planar (inflation) coordinates, which covers only half of the $dS$,
\bge
\begin{aligned}
  &X^0=-\sinh(H t)-\FR{1}{2}x_i x^i e^{H t},
 &&X^i=x^i e^{H t}, X^D=\cosh(H t)-\FR{1}{2}x_i x^i e^{H t}.
\end{aligned}
\ede
The metric written in inflation coordinates is the familiar one in inflation calculation,
\bge
  \di s^2=-\di t^2+e^{2Ht}\di x^i\di x^i.
\ede
The conformal time $\tau$ used in this paper is related to $t$ via $e^{Ht}=-1/(H\tau)$, and the dS metric expressed in $(\tau,x^i)$ coordinates is given by (\ref{dSc}). 

The geodesic distance $L(x,x')$ between two points $x,x'$ on sphere is simply proportional to the angle between the corresponding vectors $\vec X$ and $\vec X'$ in background Euclidean space. Let this angle be $\theta$, then we have $L(x,x')=\theta/H$. Meanwhile we will also use another spherically invariant distance between $x$ and $x'$, i.e. the imbedding distance $Z(x,x')\equiv H^{2}\vec X\cdot \vec X'=\cos(HL(x,x'))$. It is useful to write the imbedding distance $Z(x,x')$ in conformal coordinates $x=(\tau,\mb x)$, $x'=(\tau',\mb x')$, as,
\bge
\label{Zinf}
Z(x,x')=1-\FR{|\mb x-\mb x'|^2-(\tau-\tau')^2}{2\tau\tau'}.
\ede

Given a function of geodesic distance $f=f(L)$ or imbedding distance $f=f(Z)$, we need to know how to take derivate of it. For this purpose we only need to know that the covariant derivative of the geodesic distance itself is the unit normal vector tangent to it, which we denote as $n_\mu$ and $n_{\mu'}$ for the two ends of $L(x,x')$, respectively,
\begin{align}
  &n_\mu\equiv\nabla_\mu L(x,x'),
  &&n_{\mu'}\equiv\nabla_{\mu'} L(x,x').
\end{align}
Then on sphere these vectors satisfy $n^\mu n_\mu=1$, $n^{\mu'}n_{\mu'}=1$. The covariant derivatives of these vectors can be further represented in terms of themselves, as,
\begin{align}
  \nabla_\mu n_\nu=&~H\cot(HL)(g_{\mu\nu}-n_\mu n_\nu),\\
  \nabla_\mu n_{\nu'}=&-H\csc(HL)(g_{\mu\nu'}+n_\mu n_{\nu'}),\\
  \nabla_\rh g_{\mu\nu'}=&~H\tan(HL/2)(g_{\rh\mu}n_{\nu'}+g_{\rh\nu'} n_{\mu}).
\end{align}
For functions of imbedding distance $Z(x,x')$ we can also get similar expressions, by noting that $\di Z/\di L=-H\sqrt{1-Z^2}$. We refer readers to \cite{Synge1960,Allen1986} for more details on concepts and quantities in this paragraph.

\subsection{Spherical Harmonics}

The spherical harmonics on $S^D$ are defined to be eigenfunctions of Laplacian operator \cite{Higuc1987,Marolf1006},
\bge
\nabla^2 Y_{\vec L}(x)=-H^2 L(L+d)Y_{\vec L}(x).
\ede
$Y_{\vec L}$ is parameterized by a vector $\vec L=(L_D,\cdots, L_1)$ with all $L_i~(i=1,\cdots,D)$ being integers and satisfying $L_D\geq\cdots\geq L_2\geq |L_1|$. In above expression and thought the paper we also use the notation $L=L_D$. The spherical harmonics satisfy the following orthonormal conditions,
\bge
\begin{aligned}
  &\sum_{\vec L}Y_{\vec L}^{}(x)Y_{\vec L}^*(x')=H^{-D}\delta(x,x'),
  &&\int\di\Omega\, Y_{\vec L}^{}(x)Y_{\vec M}^*(x)=H^{-D}\de_{\vec L\vec M}.
\end{aligned}
\ede
The integral measure $\di\Omega\equiv\di^Dx\sqrt{g(x)}$ and the delta function are defined in the covariant way,
\bge
  \int\di\Omega\, \de(x,x')f(x)=f(x').
\ede
We also use the shorthand notation $\di\Omega'\equiv\di^Dx'\sqrt{g(x')}$ occasionally.

When dealing with spinors we also make use of spin-weighted spherical harmonics \cite{Drumm1979}, which are defined to be the eigenfunctions of spinor rotation generator $J_S=\frac{1}{8}[\ga^M\ga^N](X_M\pd_N-X_N\pd_M)$ of the background $(D+1)$-dimensional Euclidean space acting on the sphere,
\bge
\label{SpSH}
\begin{aligned}
  &(J_S+\FR{D}{2})Y_{\vec L s}^{\pm}(x)=\lam_{L}^\pm Y_{\vec L s}^\pm(x),
  &&\lam_L^\pm=\mp (L+\FR{d\pm 1}{2}),
\end{aligned}
\ede
where $s$ is the spinor index. Here the additional constant $D/2$ is added to $J_S$ because the Dirac operator $\sla\nabla$ on sphere can be rewritten as $\sla\nabla=H^{2}\sla X(J_S+D/2)$ where the background coordinates $X^i$ is subject to the restriction $H^2X^2=1$.

The spin-weighted spherical harmonics can be constructed from ordinary $Y_{\vec L}(x)$'s and a set of basis of constant Dirac spinor $\psi_s$ by projections,
\bge
\begin{aligned}
  &Y_{\vec L s}^\pm (x)= P_\pm Y_{\vec L}(x)\psi_s,
  &&P_+=\FR{L+d-J_S}{2L+d}, &&P_-=1-P_+.
\end{aligned}
\ede
The spin-weighted spherical harmonics satisfy the following relations,
\bgs
\label{SpSHON}
\begin{align}
 &\int\di\Omega\, Y_{\vec L s}^{\pm\dag}(x)Y_{\vec M s'}^{\pm}(x)=H^{-D}\de_{\vec L\vec M}\de_{ss'},\\
 &\int\di\Omega\, Y_{\vec L s}^{\pm\dag}(x)Y_{\vec M s'}^{\mp}(x)=0,\\
 &\sum_{\vec L, s}\Big[Y_{\vec L s}^+(x)Y_{\vec L s}^{+\dag}(x')+Y_{\vec L s}^-(x)Y_{\vec L s}^{-\dag}(x')\Big]=H^{-D}\de(x,x')\mb{1},
\end{align}
\eds
where $\mb{1}$ stands for unit spinor. By definition, we can rewrite a scalar spherical harmonic function $Y_{\vec L}(x)$ in terms of spin-weighted one, as,
\bge
\label{SpSHDecomp}
  Y_{\vec L}(x)\mb{1}=\sum_s\Big[Y_{\vec Ls}^+(x)+Y_{\vec Ls}^-(x)\Big]\psi_s^\dag.
\ede

\subsection{Propagators}

The propagator for a real scalar field of mass $M$ is given by,
\bge
\label{ScaPro}
  G(x,x')=\FR{H^{D-2}}{(4\pi)^{D/2}}\FR{\Gamma(d/2-\mu)\Gamma(d/2+\mu)}{\Gamma(d/2)}\;{}_2F_1\Big(\FR{d}{2}-\mu,\FR{d}{2}+\mu;\FR{D}{2};\FR{1+Z_{xx'}}{2}\Big),
\ede
where $\mu\equiv\sqrt{d^2/4-(M/H)^2}$. It can also be expressed in terms of spherical harmonics, as,
\bge
\label{ScaProSH}
\begin{aligned}
&G(x,x')=H^{D-2}\sum_{\vec L}\FR{1}{\lam_L}Y_{\vec L}(x)Y_{\vec L}^*(x'),
&&\lam_L=(L+\FR{d}{2}-\mu)(L+\FR{d}{2}+\mu).
\end{aligned}
\ede

In the Euclidean dS calculations in Sec. 2 we shall need the propagator for \emph{massless} Dirac spinor only \cite{Drumm1979}. It can be conveniently represented in terms of spinor spherical harmonics as,
\begin{align}
  G_F(x,x')=&~H^D\sla X\sum_{\vec L,s}\bigg[\FR{1}{\lam_L^+}Y_{\vec L s}^+(x)Y_{\vec L s}^{+\dag}(x')+\FR{1}{\lam_L^-}Y_{\vec L s}^-(x)Y_{\vec L s}^{-\dag}(x')\bigg],\n\\
   =&-H^D\sum_{\vec L,s}\bigg[\FR{1}{\lam_L^+}Y_{\vec L s}^+(x)Y_{\vec L s}^{+\dag}(x')+\FR{1}{\lam_L^-}Y_{\vec L s}^-(x)Y_{\vec L s}^{-\dag}(x')\bigg]\sla X',
\end{align}
where $\lam_L^\pm$ is defined in (\ref{SpSH}).

We shall also need the propagator for a \emph{massive} vector field of mass $M_A$ \cite{Allen1986},
\begin{align}
\label{VP}
  G_{\mu\nu'}(x,x')=&~\FR{(1-D)H^D}{2(4\pi)^{D/2}M_A^2}\bigg[\Big(\FR{1-Z^2}{D-1}\FR{\di}{\di Z}+Z\Big)(g_{\mu\nu'}+n_\mu n_{\nu'})-n_{\mu}n_{\nu'}\bigg]\n\\
  &~\times\FR{\Gamma(\frac{D+1}{2}+\mu_1)\Gamma(\frac{D+1}{2}-\mu_1)}{\Gamma(\frac{D}{2}+1)}\n\\
  &~\times{\,}_2F_1\Big(\FR{D+1}{2}+\mu_1,\FR{D+1}{2}-\mu_1;\FR{D}{2}+1;\FR{1+Z}{2}\Big),
\end{align}
where $\mu_1\equiv \sqrt{(D-3)^2/4-(M_A/H)^2}$, and $Z=Z(x,x')$ is again the imbedding distance between $x$ and $x'$. It is interesting to note that the propagator above is inversely proportional to $M_A^2$ and is divergent as $M_A^2\to 0$, which is similar to the case in flat spacetime. This means in particular that the massless propagator for the gauge boson is not the massless limit of the massive propagator. In fact, the massless propagator for gauge boson is more complicated (partly due to gauge freedom) and we refer the readers to \cite{Allen1986} for details.

\subsection{Simple Manipulations}

One simple but very useful relation which is particularly manifest in Euclidean dS is the following,
\bge
\label{GGInt}
  \int\di\Omega_x G_1(x_1,x)G_2(x,x_2)=-\FR{G_1(x_1,x_2)-G_2(x_1,x_2)}{m_1^2-m_2^2},
\ede
where $G_{1,2}(x,x')$ denotes a scalar propagator with mass $m_{1,2}$. The proof of this relation is straightforward,
\begin{align}
  &\int\di\Omega_x\, G_1(x_1,x)G_2(x,x_2)=\sum_{\vec L,\vec M}\FR{H^{2D-4}}{\lam_{1L}\lam_{2M}}Y_{\vec L}(x_1)Y_{\vec M}^*(x_2)\int\di\Omega_x Y_{\vec L}^*(x)Y_{\vec M}(x)\n\\
  =&\sum_{\vec L}\FR{H^{D-4}}{\lam_{1L}\lam_{2L}}Y_{\vec L}(x_1)Y_{\vec L}^*(x_2)=\FR{-H^{D-2}}{m_1^2-m_2^2}\sum_{\vec L}\Big(\FR{1}{\lam_{1L}}-\FR{1}{\lam_{2L}}\Big)Y_{\vec L}(x_1)Y_{\vec L}^*(x_2)\n\\
  =&-\FR{G_1(x_1,x_2)-G_2(x_1,x_2)}{m_1^2-m_2^2}.
\end{align}
In particular, in the limiting case when $m_1=m_2$, we have,
\bge
  \int\di\Omega_x\,G_1(x_1,x)G_1(x,x_2)=-\FR{\pd}{\pd m^2}G_m(x_1,x_2)\Big|_{m=m_1},
\ede
and one can even generalize it to the product of a string of propagators, and the result is actually put in use (See Eq. (5.5) of \cite{Chen:2016nrs} and the discussion nearby) when we demonstrate the equivalence between dynamical renormalization group resummation and the explicit summation of all mass insertions to all orders in perturbation theory.

Actually the relation (\ref{GGInt}) is nothing but the leading order of the perturbation expansion for a bilinear mixing between to scalar fields of masses $m_1$ and $m_2$, and one can actually check that the relation holds using the standard in-in formulation in real-time dS,\begin{align}
&-\ii\int^\tau_{-\infty}\FR{\di\tau'}{(H\tau')^4}\,\Big[G_{++}^{(\chi)}(\mb k,\tau,\tau')G_{++}^{(\phi)}(\mb k,\tau',\tau)-G_{+-}^{(\chi)}(\mb k,\tau,\tau')G_{-+}^{(\phi)}(\mb k,\tau',\tau)\Big]\n\\
=&-\ii\int^\tau_{-\infty}\FR{\di\tau'}{(H\tau')^4}\,\Big[\chi_k(\tau)\chi_k^*(\tau')\phi_k^*(\tau')\phi_k(\tau)-\text{c.c.}\Big]\n\\
=&~\FR{-\ii\pi^2}{16}\bigg[H_{\nu_\chi}^{(1)}(-k\tau)H_{\nu_\phi}^{(1)}(-k\tau)\int_{-\infty}^{\tau}\FR{\di\tau'}{(-\tau')^3}H_{\nu_\chi}^{(1)*}(-k\tau)H_{\nu_\phi}^{(1)*}(-k\tau)-\text{c.c.}\bigg]\n\\
=&~\FR{-\pi}{4(\nu_\chi^2-\nu_\phi^2)}\Big[\big| H_{\nu_\chi}^{(1)}(-k\tau)\big|^2-\big| H_{\nu_\phi}^{(1)}(-k\tau)\big|^2\Big]\n\\
=&~\FR{1}{M_\chi^2-M_\phi^2}\Big[G_{++}^{(\chi)}(\mb k,\tau,\tau)-G_{++}^{(\phi)}(\mb k,\tau,\tau)\Big],
\end{align}
in which we have used $\phi$ and $\chi$ to denote the two scalar fields with mass $M_\chi$ and $M_\phi$. Although this relation may look trivial, it can bring significant simplifications in loop calculation, as shown in several examples in Sec.\;\ref{Sec_2pt}.

\end{appendix}

}

\end{document}